\documentclass[onecolumn,epjc3]{svjour3} 
 
\usepackage[T1]{fontenc}
\usepackage{epsfig,amsmath,amssymb}
\usepackage{graphicx}
\usepackage{xspace}
\usepackage{listings}
\usepackage{ulem}
\usepackage{slashed}
\usepackage{cite}
\usepackage{nicefrac}
\usepackage{feynmp}
\usepackage[colorlinks,citecolor=blue,urlcolor=blue,linkcolor=blue]{hyperref}
\journalname{Eur. Phys. J. C}

\newcommand\FH{{\tt FeynHiggs}\xspace}
\newcommand\CPH{{\tt CPsuperH}\xspace}

\newcommand\SARAH{{\tt SARAH}\xspace}

\newcommand\SPheno{{\tt SPheno}\xspace}

\newcommand{\DRbar}{{\ensuremath{\overline{\mathrm{DR}}}}\xspace}

\newcommand{\GeV}{{\ensuremath{\text{GeV}}}}
\newcommand{\TeV}{{\ensuremath{\text{TeV}}}}

\newcommand\NC{{\tt NMSSMCALC}\xspace}

\newcommand{\newc}{\newcommand}

\newc{\Psibar}{\overline{\Psi}}
\newc{\FFbS}{\overline{FF}S}
\newc{\FFbV}{\overline{FF}V}
\newc{\FFbe}{\overline{FF}\epsilon}
\newc{\FSS}{F_{SS}}
\newc{\FSSS}{F_{SSS}}
\newc{\FFFS}{F_{FFS}}
\newc{\FFFbS}{F_{\overline{FF}S}}
\newc{\FSSV}{F_{SSV}}
\newc{\FVS}{F_{VS}}
\newc{\FVVS}{F_{VVS}}
\newc{\FFFV}{F_{FFV}}
\newc{\FFFbV}{F_{\overline{FF}V}}
\newc{\FVV}{F_{VV}}
\newc{\FVVV}{F_{VVV}}
\newc{\fggV}{f_{ggV}}
\newc{\Fgauge}{F_{\rm gauge}}
\newc{\fSS}{f_{SS}}
\newc{\fSSS}{f_{SSS}}
\newc{\fFFS}{f_{FFS}}
\newc{\fFFbS}{f_{\overline{FF}S}}
\newc{\fSSV}{f_{SSV}}
\newc{\fVVS}{f_{VVS}}
\newc{\fVS}{f_{VS}}
\newc{\fFFV}{f_{FFV}}
\newc{\fFFbV}{f_{\overline{FF}V}}
\newc{\fVV}{f_{VV}}
\newc{\fVVV}{f_{VVV}}
\newc{\fgauge}{f_{\rm gauge}}
\newc{\FFVbar}{{\overline{FF}V}}
\newc{\FFSbar}{\overline{FF}S}

\newcommand{\nn}{\nonumber}

\newc{\bsigmu}{\bar\sigma^\mu}

\def\twov[#1,#2]{\left( \begin{array}{c} #1 \\ #2 \end{array} \right)}
\def\twoa[#1,#2][#3,#4]{\left( \begin{array}{cc} #1 & #2 \\ #3 & #4 \end{array} \right)}
\def\ov{\overline}

\begin{document}

\title{The Higgs mass in the CP violating MSSM, NMSSM, and beyond}
\author{
   Mark D. Goodsell \thanksref{a1} \and
   Florian Staub \thanksref{a2}
   }

\institute{ Sorbonne Universit\'es, UPMC Univ Paris 06, UMR 7589, LPTHE, F-75005, Paris, \& \\
CNRS, UMR 7589, LPTHE, F-75005, Paris, France\label{a1}
\and
Theory Department, CERN, 1211 Geneva 23, Switzerland\label{a2}
}

\date{CERN-TH-2016-089}

\maketitle

\lstset{frame=shadowbox}
\lstset{prebreak=\raisebox{0ex}[0ex][0ex]
        {\ensuremath{\hookrightarrow}}}
\lstset{postbreak=\raisebox{0ex}[0ex][0ex]
        {\ensuremath{\hookleftarrow\space}}}
\lstset{breaklines=true, breakatwhitespace=true}
\lstset{numbers=left, numberstyle=\scriptsize}
 
\begin{abstract}
We discuss the automatised calculation of the Higgs mass in renormalisable supersymmetric models with complex parameters at the two-loop level. Our setup is based on
the public codes \SARAH and \SPheno, which can now compute the two-loop corrections to masses of all neutral scalars in such theories. The generic ansatz for these calculations and the handling of the `Goldstone Boson catastrophe' is described. It is shown that we find perfect 
agreement with other existing two-loop calculations performed in the \DRbar scheme. We also use the functionality to derive results for the MSSM and NMSSM not available before: the Higgs mass in the constrained version of the complex MSSM, and the impact of CP phases in the two-loop corrections beyond $O(\alpha_s \alpha_t)$ for the scale invariant NMSSM are briefly analysed.
\end{abstract}


\section{Introduction}
The discovery of the Higgs boson was the biggest success of Run-I of the Large Hadron Collider (LHC) \cite{Chatrchyan:2012ufa,Aad:2012tfa}. The mass of the Higgs is already known very precisely up to a few hundred MeV and its properties are in good agreement with the expectations from the Standard Model (SM). These observations now give strong constraints on any extension of the SM. Therefore, it is necessary to calculate these properties with increasing accuracy to close the gap between the experimental and theoretical uncertainty. In the context of supersymmetric models, most effort has been put into a precise calculation of the Higgs mass in the minimal supersymmetric standard model (MSSM) assuming real parameters\cite{Haber:1990aw,Ellis:1990nz,Okada:1990vk,Okada:1990gg,Ellis:1991zd,Brignole:1992uf,Chankowski:1991md,Dabelstein:1994hb,Pierce:1996zz,Hempfling:1993qq,Carena:1995wu,Heinemeyer:1998jw,Zhang:1998bm,Heinemeyer:1998np,Heinemeyer:1999be,Espinosa:1999zm,Espinosa:2000df,Brignole:2001jy,Degrassi:2001yf,Martin:2002wn,Brignole:2002bz,Dedes:2002dy,Martin:2002iu,Dedes:2003km,Heinemeyer:2004xw,Sasaki:1991qu,Carena:1995bx,Haber:1996fp,Espinosa:2001mm,Martin:2007pg,Kant:2010tf,Harlander:2008ju,Martin:2004kr,Borowka:2014wla,Hollik:2014bua,Degrassi:2014pfa}. In addition, for the next-to-minimal supersymmetric standard model (NMSSM) most calculations of the Higgs mass considered the CP conserving case \cite{Degrassi:2009yq,Staub:2010ty,Ender:2011qh,Goodsell:2014pla}. 

Recently, however, there has been increasing interest in theories that extend these minimal models, and this has led to the present authors' work\cite{Goodsell:2014bna,Goodsell:2015ira}, building on that of \cite{Martin:2001vx,Martin:2003qz,Martin:2003it,Martin:2005eg}, to extend those calculations and provide a public implementation for two-loop Higgs mass calculations in generic theories. Up to this point the corrections were only available for CP-even scalars in theories with real parameters; this paper discusses the extension to all neutral scalars with or without CP violation (CPV). 

Indeed, the focus on the real versions of the MSSM and NMSSM to study the Higgs mass can hardly be motivated from first principles: there is no strong argument why the CP phases in the soft-breaking sector of SUSY models should be small -- especially if SUSY breaking is transmitted via gravity. Moreover, SUSY models with CPV can have very interesting phenomenological aspects, see for instance \cite{Carena:2002bb,Bartl:2006yv,Williams:2007dc,Chiang:2008ud,Terwort:2012sy,Bharucha:2013epa,Munir:2013dya,Heinemeyer:2015qbu,Moretti:2015bua,Carena:2015uoe}. Putting the phases to zero is often an assumption to circumvent conflicts with experimental limits and simplify calculations.  Due to this the impact of CP phases on the Higgs mass has so far only been partially considered in both models. In the MSSM it was first studied using renormalisation group  techniques \cite{Carena:2000dp,Carena:2000yi,Carena:2001fw}, while diagrammatic calculations at the one-loop level \cite{Frank:2006yh} and particular two-loop level \cite{Hollik:2014wea,Hollik:2014bua} were performed much later. For the complex NMSSM the one-loop results \cite{Graf:2012hh} are so far only accompanied by two-loop corrections of $O(\alpha_s \alpha_t)$ \cite{Muhlleitner:2014vsa}. \\

CPV in supersymmetric theories is constrained by several observations, principally meson mixing (in particular $K^0$, $B^0_{s}, B^0_d$ and $D^0$ mesons) and decays; the electric dipole moment of nucleons and electrons; and Higgs coupling measurements. Meson physics typically places extremely stringent bounds, but there is little overlap between those constraints and constraints on the Higgs mass/mixings as relevant for this work, because generational mixing is required -- and even then, large enough generational mixing to have a sizeable effect on the Higgs mass at two loops may be unconstrained by flavour \cite{Goodsell:2015yca}. Furthermore, the measurement of the Higgs couplings at the LHC is rather insensitive to parity violation, with the parity-violating couplings still allowed to be of the same order as, if somewhat less than\footnote{The $0^-$ hypothesis for the Higgs boson is excluded, but couplings up to $0.83$ times the Standard-Model values are still allowed for parity-violating couplings to Z-bosons of the form $S Z_{\mu \nu}\tilde{Z}^{\mu \nu}$.}, the Standard-Model-like ones \cite{ATLAS:2015zja}, and so direct searches for additional Higgs bosons actually place more stringent constraints. 

Therefore the most relevant constraint on the parameter space that we shall consider comes from electric dipole moments, in particular that of the electron $d_e$, which is constrained to be \cite{Baron:2013eja}
\begin{align}
|d_e| < 8.9 \times 10^{-29}\ e\ \mathrm{cm} = 4.5\times 10^{-15} e\ \mathrm{GeV}^{-1}.
\end{align}
The typical value for electric or chromoelectric moments for fermions $i$ of mass $m_i$ and a common SUSY scale $M_{\rm SUSY}$ is \cite{Pospelov:2005pr} 
\begin{align}
\kappa_i \equiv \frac{m_i}{16\pi^2 M_{\rm SUSY}^2} = 1.3 \times 10^{-25}\ {\rm cm} \times \frac{m_i}{\rm MeV} \times \left( \frac{{\rm TeV}}{M_{\rm SUSY}} \right)^2 
\end{align}
multiplied by a numerical factor, three Yukawa or gauge couplings, and the sine of a CP-violating phase. 
In the case of the electron dipole moment in the MSSM with only CP-violation entering through the $\mu$-term we have 
\begin{align}
d_e/e \simeq & \frac{5 g^2 }{24} \kappa_e \tan \beta \sin ( \varphi_\mu)
\end{align}
and we therefore need a large suppression of the total angle $\varphi_\mu$ by roughly three orders of magnitude; however, if we just consider the sneutrino-chargino sector then we obtain in the same limit 
\begin{align}
|d_e/e| \simeq & \frac{4 g^2 }{3} \kappa_e \tan \beta \big|\sin ( \varphi_\mu + \varphi_{M_2} + \eta)\big|
\label{EQ:EDMchargino}\end{align}
which similiarly constrains more CP-violating phases.

There is a constraint from the neutron electric dipole moments, a recent limit being \cite{Afach:2015sja} (see also \cite{Baker:2006ts,Baker:2007df,Serebrov:2015idv})
\begin{align}
|d_n/e| \lesssim& 3.0 \times 10^{-26}\ \mathrm{cm} .
\end{align}
Here the calculation is more complicated, since it depends on the electric dipole moments of the light squarks, their chromoelectric moments $\tilde{d}_{u,d}$ (naively suppressed by a factor $\frac{e}{4\pi}$), and the theta-angle of QCD. While this has the power to restrict the gluino phase through a squark-gluino loop, either through a direct EDM since the squarks are charged, or through the chromoelectric moments, this is not relevant for our study because the bound is not sufficiently strong: it can be easily satisfied just by, for example, taking the first two generations of squarks to have masses of a few TeV.

There is an additional strong constraint from the mercury dipole moment\cite{Pospelov:2005pr}:
\begin{align}
|d_{\rm Hg}/e| \simeq& 10^{-28} \ \mathrm{cm}\ \times \left| \frac{\tilde{d}_u - \tilde{d}_d}{10^{-25}\ \mathrm{cm}} \right| \lesssim  2\times 10^{-28} \ \mathrm{cm} .
\end{align}
If the first two generations of squarks are heavy then, again, this will not constrain the parameter space relevant for the Higgs mass. 

In summary, in supersymmetric models some sources of CPV in the Higgs sector are required to be small by experiment, but several parameters are essentially unconstrained -- which could have a strong impact on the masses of the neutral and charged scalars. In general, it is the electric dipole moment of the electron that will restrict the phases  $\varphi_\mu, \eta$ and the phases of the electroweakinos to be close to zero, so we will not consider their effect on the Higgs masses; on the other hand, we shall treat the phase of the gluino and trilinears in the third generation of squarks to be important free parameters, keeping the first two generations of squarks heavy.

The aim of this work is to present the possibility of calculating the Higgs mass and that of all other neutral scalars in a wide range of supersymmetric models with and without CPV to the same accuracy: an automatised, diagrammatic calculation of the Higgs mass covering CPV at the two-loop level is now available via the combination of the public codes \SARAH \cite{Staub:2008uz,Staub:2009bi,Staub:2010jh,Staub:2012pb,Staub:2013tta,Staub:2015kfa} and \SPheno \cite{Porod:2011nf,Porod:2003um}. This functionality extends the automatised two-loop calculations for the real case presented in Refs.~\cite{Goodsell:2014bna,Goodsell:2015ira}. In general, the calculations are done in the gaugeless limit and neglecting the dependence of the external momenta, i.e. they are competitive with the current state-of-the-art calculations for the complex MSSM, but extend any existing two-loop calculation for other SUSY models by important corrections beyond $O(\alpha_s \alpha_t)$. We explain in sec.~\ref{SEC:METHOD} the underlying methodology used in the calculations and some technical subtleties of the new extension before we present in sec.~\ref{SEC:VALIDATION} the validation of the routines in the presence of complex parameters. In secs.~\ref{SEC:MSSM} and \ref{SEC:NMSSM} we discuss some applications of these routines in the context of the MSSM and NMSSM, before we conlcude in sec.~\ref{sec:conclusion}.

\section{Methodology}
\label{SEC:METHOD}

The calculation of CP-violating corrections at two loops is now available in \SARAH via the diagrammatic approach described in Ref.~\cite{Goodsell:2015ira}. Indeed, no modifications are required to the expressions given in that paper. For the computation of masses for CP-odd scalars in CP-conserving theories the same routines also apply; since the formalism in Ref.~\cite{Goodsell:2015ira} is given in terms of real scalars, and the CP-odd scalars are just CP-even scalars with different labels.
However, once we extend our computations to these cases we find two potential subtleties associated with our method of avoiding the Goldstone Boson Catastrophe. 

To remind the reader, this problem highlighted and resolved in Refs.~\cite{Martin:2002wn,Martin:2014bca,Elias-Miro:2014pca} arises either in the MSSM beyond the gaugeless limit, or in theories beyond the MSSM even in the gaugeless limit, in that the \DRbar mass of goldstone bosons have indeterminate. The full on-shell mass of course being zero, the \DRbar goldstone boson mass parameter is thus of the same order as loop corrections and is small. The problem is that this parameter appears in the loop corrections to the tadpoles and masses of Higgs bosons (and other particles) and the solution for a mass calculation is to include momentum dependence. 

However, since this is computationally onerous, our solution (described in Refs.~\cite{Goodsell:2014bna,Goodsell:2014pla}) is to exploit the fact that we work in the gaugeless limit and, in our two loop calculation, can therefore neglect corrections to the mass proportional to electroweak gauge couplings: we use the full potential $V_0 + V_1 + V_2|_{\rm gaugeless}$ to solve the tadpole equations, and then use the parameters determined from these in our pure gaugeless tree-level potential $V_0|_{\rm gaugeless}$ to determine the masses in our theory. Since we are effectively working in a false minimum the \DRbar goldstone masses entering in our two-loop calculation are non-zero and of order the electroweak boson masses. This has the effect of taming the problem for most parts of the parameter space. 

The first subtlety related to this approach as concerns CP violation is that the goldstone masses are typically tachyonic, and we retain only the real part of the loop functions. It is legitimate to ask whether the tadpole and self-energy diagrams that we compute really then correspond to the first and second derivatives of the two-loop potential  once we introduce CP violation, since the complex parts of the couplings may in principle  multiply a complex loop function. However, the terms in the potential, given in Ref.~\cite{Martin:2001vx}, all have the form 
$$\Delta^{(2)} V_{\rm (given\ topology)} = \bigg( \mathrm{Real\ product\ of\ couplings}\bigg) \times \bigg( \mathrm{loop\ function\ of\ masses}\bigg) $$
with the exception of one contribution involving fermions and scalars, given by
$$
V^{(2)}_{\ov{FF}S} = {1\over 4} y^{IJk} y^{I'J'k} M^*_{II'} M^*_{JJ'} 
                             f_{\ov{FF}S} (m^2_I, m^2_J, m^2_k) + {\rm c.c.}.
$$
However, this should be understood as $\mathrm{Re} ({1\over 2} y^{IJk} y^{I'J'k} M^*_{II'} M^*_{JJ'}) f_{\ov{FF}S} (m^2_I, m^2_J, m^2_k)$ and so falls into the same class as the other terms. Then, when we take the derivatives of the loop functions, since the masses are real, their derivatives with respect to real scalars are real, and we find that the imaginary part of the derivatives of the potential is always the same as the derivatives of the imaginary part, as we require for consistency of our approach. 

The second subtlety once we calculate the masses of CP-odd scalars, or when we have CP violation which mixes originally even and odd scalars, is that among our scalars we now have (would-be) goldstone bosons. We must therefore ensure that the final goldstone boson masses should vanish in the Landau gauge once we add the two-loop corrections to the tree and one-loop terms. 
To show that this is the case in our approach, let us write 
\begin{align}
V_0 \equiv \frac{1}{2} m_{ij}^2 S_i S_j + \tilde{V}_0 + \tilde{V}_0^D
\end{align}
for scalars $S_i$, where $\tilde{V}_0^D$ is the gauge-coupling dependent part that vanishes in the gaugeless limit. At tree level, the tadpole equations are used to determine some subset of the \DRbar mass parameters; let us take them here to be defined to be the diagonal terms:
\begin{align}
m_{0,ij}^2 v_j \equiv& - \partial_i \tilde{V}_0- \partial_i \tilde{V}_0^D \nn\\
\rightarrow m_{0,ii}^2 =& - \frac{1}{v_i} \big( \partial_i \tilde{V}_0 + \partial_i \tilde{V}_0^D \big)) - \frac{1}{v_i}\sum_{j\ne i} m_{0,ij}^2 v_j
\end{align} 
and we then compute the particle masses at tree level
\begin{align}
\mathcal{M}_{0,ij}^2 \equiv& m_{0,ij}^2 + \partial_i \partial_j \tilde{V}_0 + \partial_i \partial_j\tilde{V}_0^D \nn\\
\mathcal{M}_{0,ij}^2\big|_{\rm gaugeless} \equiv& m_{0,ij}^2 + \partial_i \partial_j \tilde{V}_0. 
\end{align}
We then compute the potential and one-loop self-energies using these \emph{tree-level} masses:
\begin{align}
\Delta V \equiv V_1 (\mathcal{M}_{0,ij}^2) + V_2 (\mathcal{M}_{0,ij}^2\big|_{\rm gaugeless}) .
\end{align}
From these we solve the tadpole corrections so that
\begin{align}
m_{ij}^2 =& m_{0,ij}^2 - \frac{\delta_{ij}}{v_i} \partial_i \Delta V \nn\\
\mathcal{M}_{ij}^2 (p^2 ) =& m_{ij}^2 + \partial_i \partial_j \tilde{V}_0 + \partial_i \partial_j\tilde{V}_0^D + \Pi_{1,ij} (p^2, \mathcal{M}_{0,ij}^2) + \partial_i \partial_jV_2 (\mathcal{M}_{0,ij}^2\big|_{\rm gaugeless}).
\end{align}
Here $ \Pi_{1,ij}$ is the one-loop self-energy.  The masses of the neutral scalars are then found as the eigenvalues of this matrix with $p^2 = m^2$ via an iterative procedure; however, for the Goldstone bosons, we merely need to verify the presence of a null eigenvector for $p^2 = 0$, when $ \Pi_{1,ij} (0, \mathcal{M}_{0,ij}^2) =  \partial_i \partial_jV_1$. For $p^2=0$ we can rewrite the above as 
\begin{align}
\mathcal{M}_{ij}^2 (0) =& \mathcal{M}_{0,ij}^2 + \Delta_1 \mathcal{M}_{ij}^2 (\mathcal{M}_{0,ij}^2) + \Delta_2 \mathcal{M}_{ij}^2 ( \mathcal{M}_{0,ij}^2\big|_{\rm gaugeless})  \nn\\
\Delta_\ell \mathcal{M}_{ij}^2 \equiv&- \frac{\delta_{ij}}{v_i} \partial_i V_\ell +  \partial_i \partial_jV_\ell .
\end{align}
To prove that our procedure retains a massless Goldstone boson, we recall the standard proof: if a potential is invariant under a global symmetry where $S_i \rightarrow S_i + \alpha_i$, then 
\begin{align}
\alpha_i  \frac{\partial V}{\partial S_i}=&0, \qquad \frac{\partial \alpha_i}{\partial S_j} \frac{\partial V}{\partial S_i} + \alpha_i\frac{\partial^2 V}{\partial S_i \partial S_j} =0.
\end{align}   
This is true order by order in perturbation theory. If we are at the minimum of the potential, then the first term in the second equation vanishes and we have a null eigenvector of the mass matrix given by $\Delta_i$. However, for the two-loop calculation we are not working at the true minimum of the potential, nor are we using the same potential; instead our tree-level potential has $\tilde{V}^D$ removed, and we solve the tadpole equations according to  
\begin{align}
\frac{\partial }{\partial S_i} \big( \frac{1}{2} \hat{m}_{ij}^2 S_i S_j + \tilde{V}_0 + \hat{V}_1 + V_2 \big) = - \frac{\partial \tilde{V}^D_0}{\partial S_i}
\end{align}
where here we have denoted by $\hat{m}_{ij}^2, \hat{V}_1$ the masses and one-loop potential, to indicate that they are in the gaugeless limit; note however that we do not compute  or require $\hat{V}_1$. 
Therefore 
\begin{align}
\alpha_i \frac{\partial^2 }{\partial S_i \partial S_j} \bigg( \frac{1}{2} \hat{m}_{ij}^2 S_i S_j + \tilde{V}_0 + \hat{V}_1 + V_2 \bigg) = - \frac{\partial \alpha_i}{\partial S_j} \frac{\partial \tilde{V}^D_0}{\partial S_i}.
\end{align}
The term on the right-hand side of this equation is necessary to give a mass to the goldstone boson at tree level to mitigate the goldstone boson catastrophe. However, all that we take from this comptuation is the derivatives of the two-loop potential; 
since we solve for the solution in the same way at tree-level and at two-loops, and because the potential $ \frac{1}{2} m_{ij}^2 S_i S_j + \tilde{V}_0$ is invariant under the global symmetry (even if the minimum we choose is not) then we find 
\begin{align}
\alpha_i \frac{\partial^2 }{\partial S_i \partial S_j} \bigg( \frac{1}{2} \hat{m}_{0,ij}^2 S_i S_j + \tilde{V}_0 \bigg) =& - \frac{\partial \alpha_i}{\partial S_j} \frac{\partial \tilde{V}^D_0}{\partial S_i} \nn\\
\rightarrow \alpha_i \big(\Delta_2 \mathcal{M}_{2,ij}^2\big) =& 0.
\end{align}
Since the one-loop computation used in actually calculating the scalar masses is performed in the minimum of the full potential it will automatically have massless goldstone bosons; then $\alpha_i \mathcal{M}_{ij}^2(0) = 0$ when including all corrections as required. It is a highly non-trivial check of our implementation that this should be true; we show this check of our code in the next section and find that it is satisfied to a high level of accuracy.

\section{Validation}
\label{SEC:VALIDATION}

\subsection{Comparison with CP-preserving case}

\begin{figure}[hbt]
\centering
\includegraphics[width=0.66\linewidth]{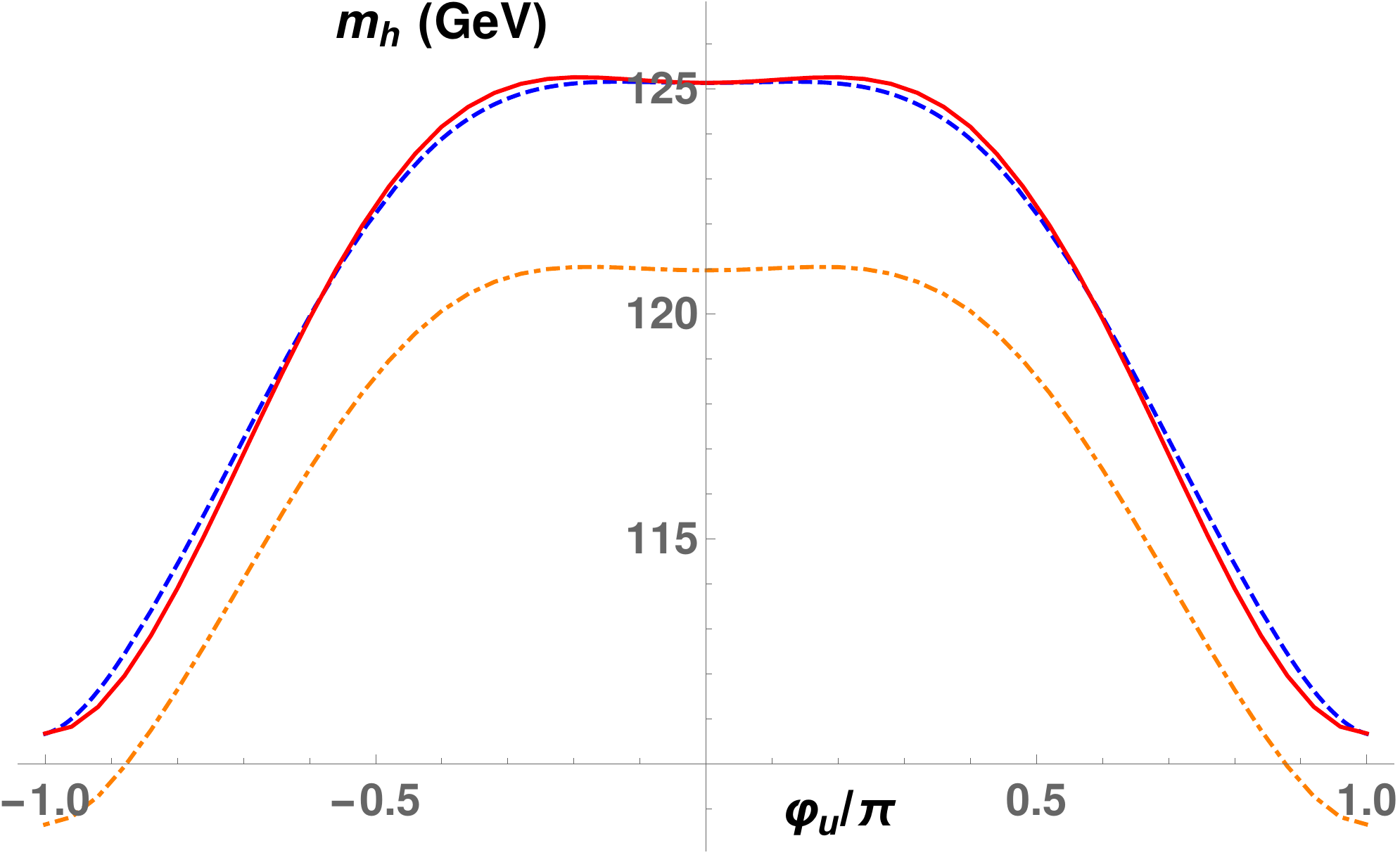} 
\caption{Plot of the lightest Higgs boson mass in the MSSM, for the parameters given in sec.~\ref{sec:MSSM_Low} with $M_3 = 2000$ \GeV, as the phase of the trilinear soft terms is varied. In red is the result of the new CP-violating code; the  blue dashed curve is a quadratic interpolation of the values $\phi_u = 0, \pi$ from the Higgs mass calculated at two loops in the CP conserving routines, where the difference is added to the one-loop CP-violating case. As can be seen we find perfect agreement for the two routines at the values $\phi_u = 0,\pm \pi$. } 
\label{fig:MSSMinter}
\end{figure}

The first verification of our new routines is to compare with the CP-conserving case. In Fig.~\ref{fig:MSSMinter} we show the one- and two-loop lightest Higgs masses obtained in our code as we vary the trilinear CP-violating phase $\varphi_u \equiv \mathrm{arg}(T_u^{3,3})$  for a point in the complex MSSM; all definitions and other parameter values are given in sec.~\ref{sec:MSSM_Low}, we take $M_3 = 2$ \TeV. On the same plot we show an \emph{interpolation} between the values in the CP-preserving MSSM for the values $\varphi_u = 0,\pi$ corresponding to $T_u^{3,3} = \pm |T_u^{3,3}$. Clearly the perfect agreement between the curves at the mid- and end-points indicates the agreement between the two codes. For the rest of the curve, the interpolation is
\begin{align}
m_h^{\rm interpolation} =& m_h^{\rm 1\ loop} (\varphi_u) + m_h^{\rm 2\ loops}\big|_{\varphi_u = 0} - m_h^{\rm 1\ loop}\big|_{\varphi_u = 0} \nn\\
& + (\varphi_u/\pi)^2 \bigg[ m_h^{\rm 2\ loops}\big|_{\varphi_u = \pi} - m_h^{\rm 1\ loop}\big|_{\varphi_u = \pi} - m_h^{\rm 2\ loops}\big|_{\varphi_u = 0} + m_h^{\rm 1\ loop}\big|_{\varphi_u = 0}\bigg]
\end{align} 
For this point, the two-loop corrections are clearly well modelled by a quadratic; we shall investigate more interesting cases in sec.~\ref{sec:MSSM_Low}.

\subsection{Check of the Goldstone mass}
\begin{figure}[hbt]
\centering
\includegraphics[width=0.66\linewidth]{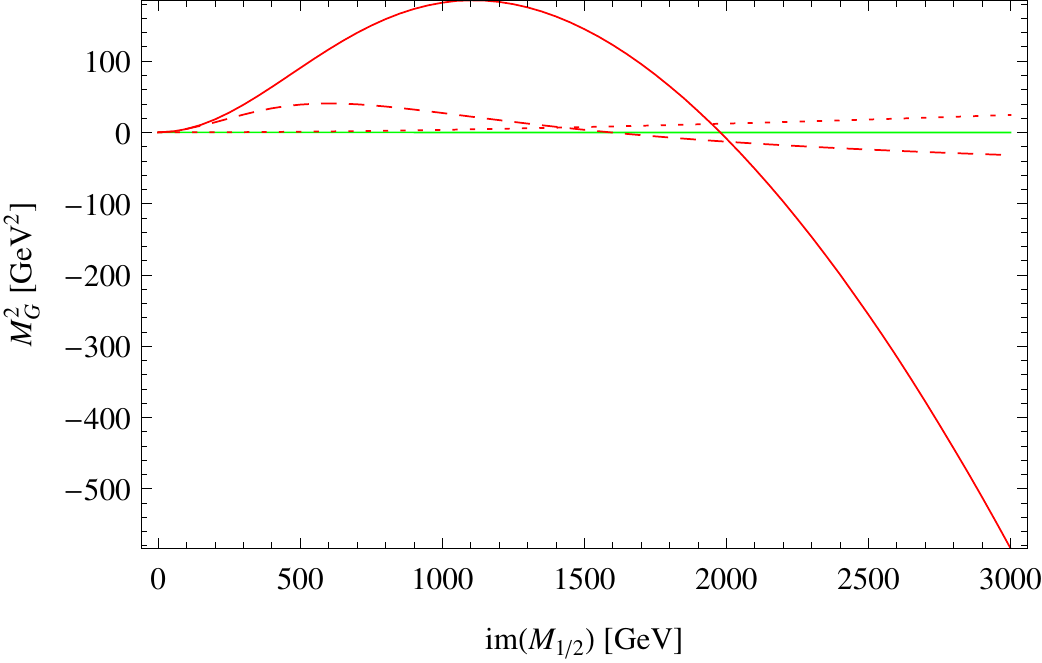} 
\caption{The calculated mass for the neutral Goldstone Boson in the MSSM in Landau gauge as function of im($M_3$). The green line corresponds to the correct calculation. 
The red lines show the impact of a possible inconsistency in the two-loop calculation: for the dashed line, the phase of $T_u$ was only included in the calculation of the masses and rotation 
matrices entering the loop calculation, but dropped in the vertices. For the full line, the phase of the rotation matrix $Z_U$ was put to zero. For the dotted line
the phase of a single vertex in a single diagram involving gluino and (s)down (!) squarks was swapped. The other input 
parameters are $m_0 = \text{re}(M_{1/2}) = 1$~TeV, $A_0 = -2$~TeV, $\tan\beta=10$, $\text{sign}(\mu)>0$.} 
\label{fig:MG}
\end{figure}

As discussed in the previous section, there is an obvious but non-trivial check for the self-consistency of the entire loop calculation: the Goldstone Boson mass has to be correct. Thus, choosing Landau gauge, the lightest 
eigenstate of the four neutral scalars must have zero mass. To obtain this in the complex case, a delicate cancellation of all phases appearing in the mass calculation of the fields in the loops, the phases in the vertices and the combination of the vertices in each diagram must happen. The impact of potential small inconsistencies in these calculations is demonstrated in Fig.~\ref{fig:MG} where we added by hand some mistakes: dropping imaginary parts of couplings only in the vertex calculation, neglecting the imaginary parts of the squark rotation matrices in the vertices, added a wrong complex conjugation to a single vertex in a single diagram. For the last point to illustrate the delicacy of the cancellations we have chosen a diagram only involving down squarks and not up squarks, which would have given an even much larger effect. While these mistakes have no impact on the results in the real case, one sees that they immediately spoil the prediction for the neutral Goldstone mass as soon as CP violation is turned on, and his provides a sensitive check for the correctness of our results.

\subsection{Comparison with the known NMSSM corrections}
\begin{figure}[tb]
\includegraphics[width=0.32\linewidth]{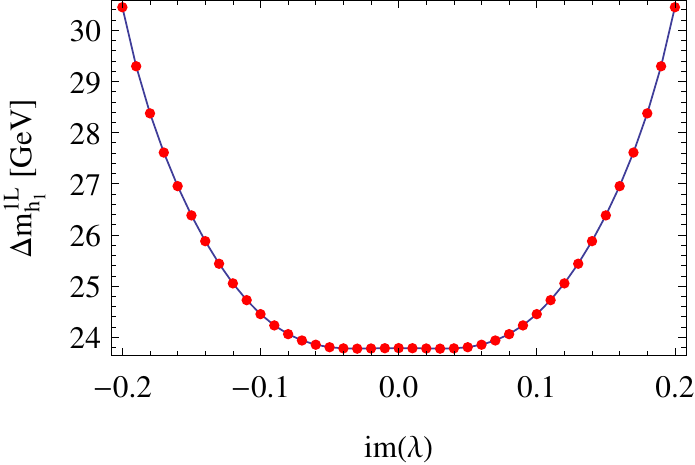}  \hfill
\includegraphics[width=0.32\linewidth]{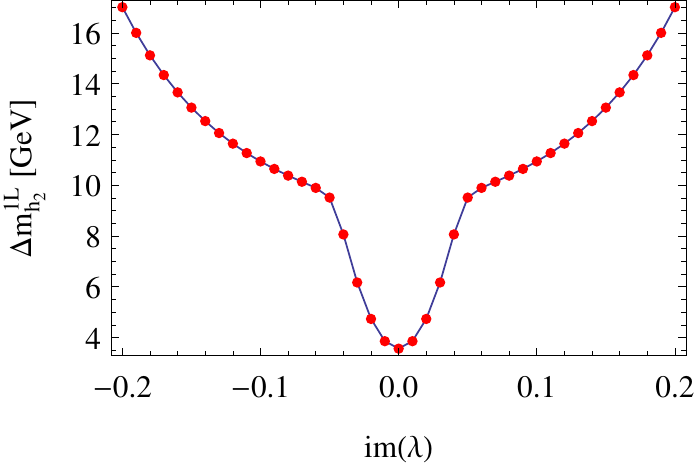}  \hfill
\includegraphics[width=0.32\linewidth]{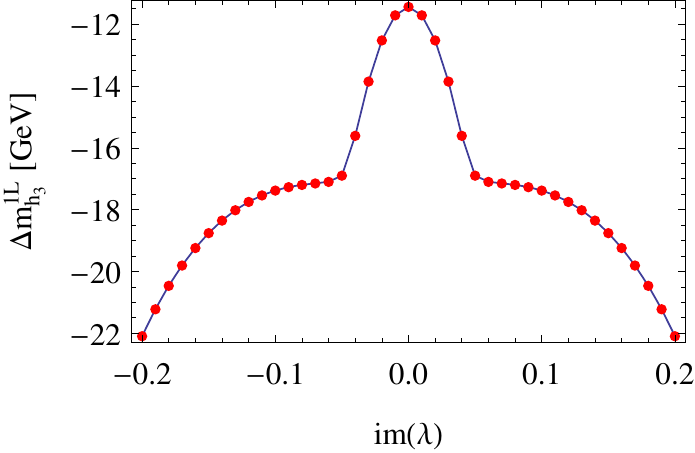}  \\
\includegraphics[width=0.32\linewidth]{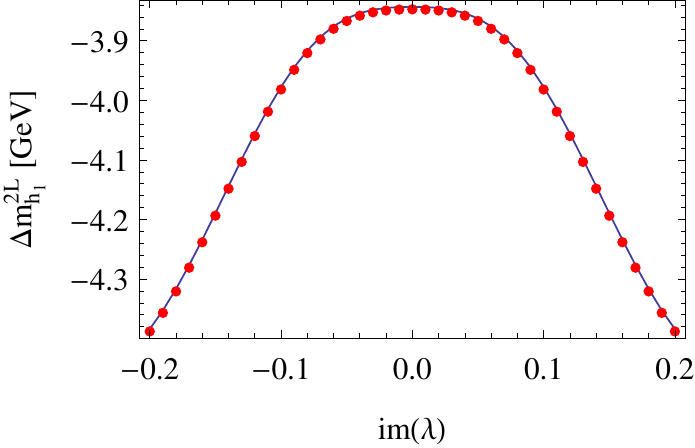}  \hfill
\includegraphics[width=0.32\linewidth]{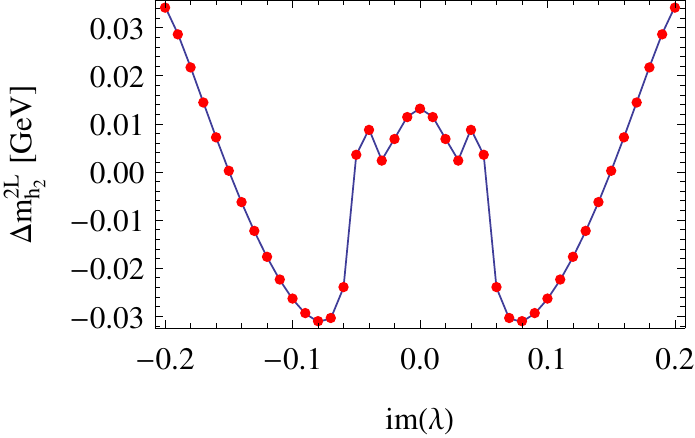}  \hfill
\includegraphics[width=0.32\linewidth]{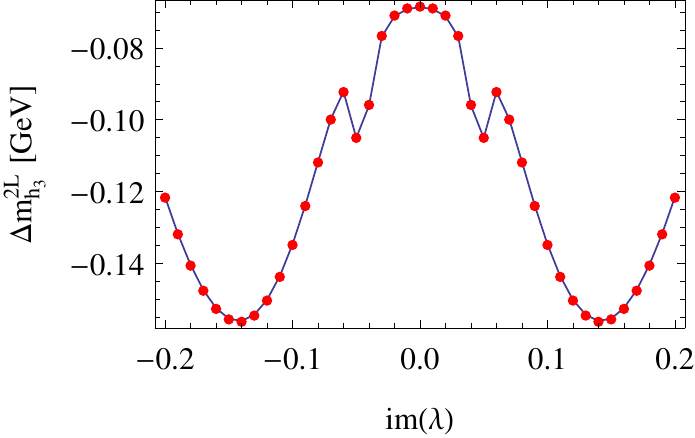}  
\caption{Effect of the one-loop corrections (first row) and two-loop corrections (second row) on the masses of the three lightest scalars as function of 
im($\lambda)$. The blue lines correspond to \SPheno, the red ones to \NC.}
\label{fig:NMSSMlam}
\end{figure}
\begin{figure}[tb]
\centering
\includegraphics[width=0.32\linewidth]{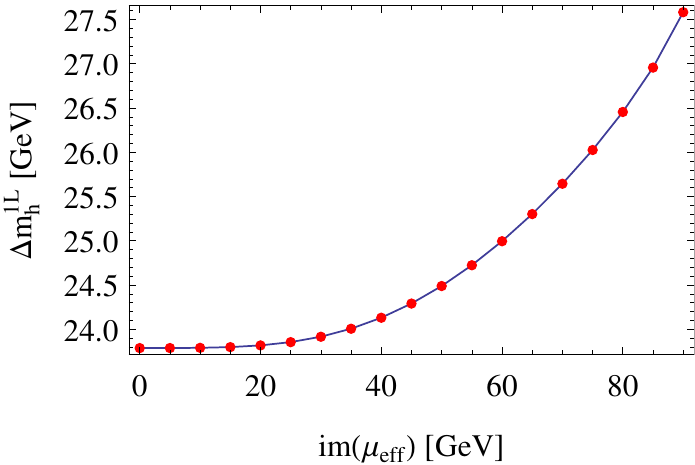}  \hspace{1cm}
\includegraphics[width=0.32\linewidth]{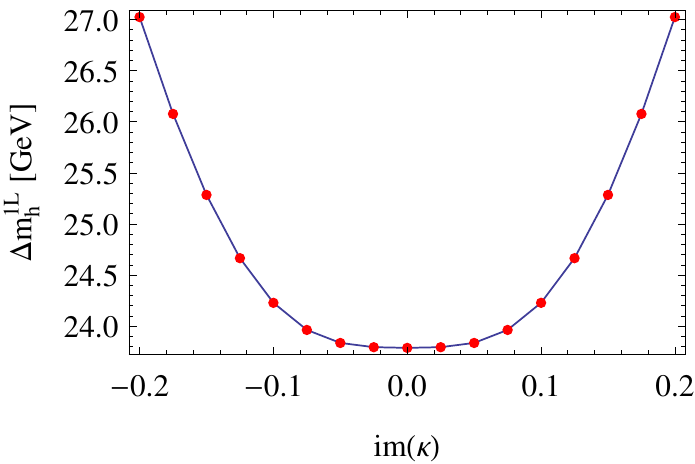}  \\
\includegraphics[width=0.32\linewidth]{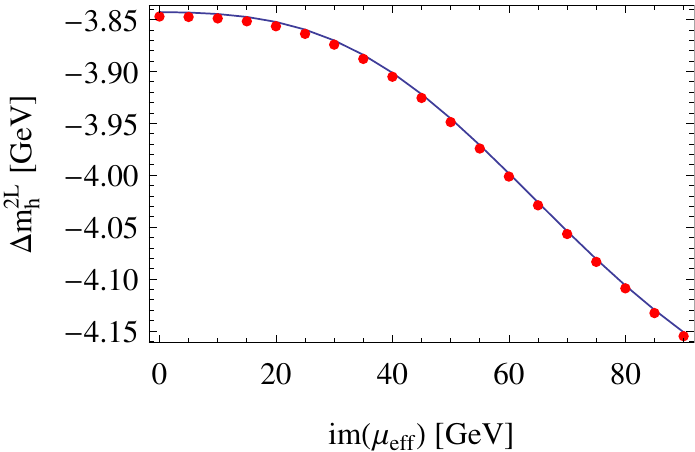}  \hspace{1cm}
\includegraphics[width=0.32\linewidth]{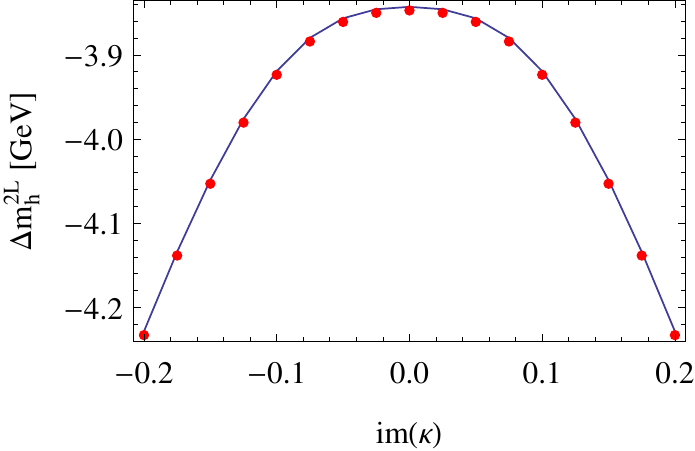}  
\caption{Effect of the one-loop corrections (first row) and two-loop corrections (second row) on the mass of the lightest scalars as function of 
im($\mu_{\rm eff}$) (left) and im($\kappa$) (right). The blue lines correspond to \SPheno, the red ones to \NC.  }
\label{fig:NMSSM2}
\end{figure}
\begin{figure}[tb]
\includegraphics[width=0.32\linewidth]{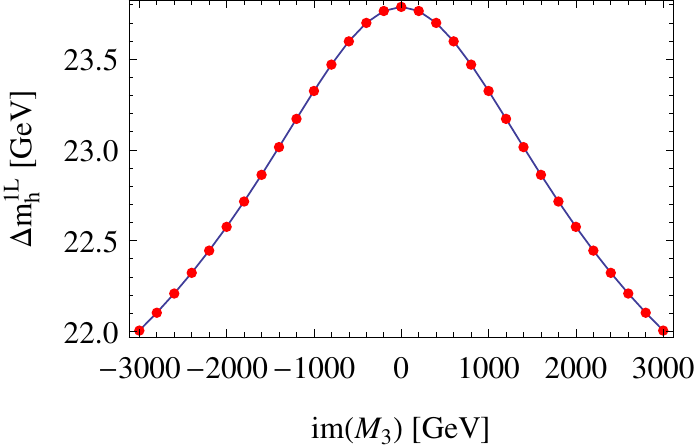}  \hfill
\includegraphics[width=0.32\linewidth]{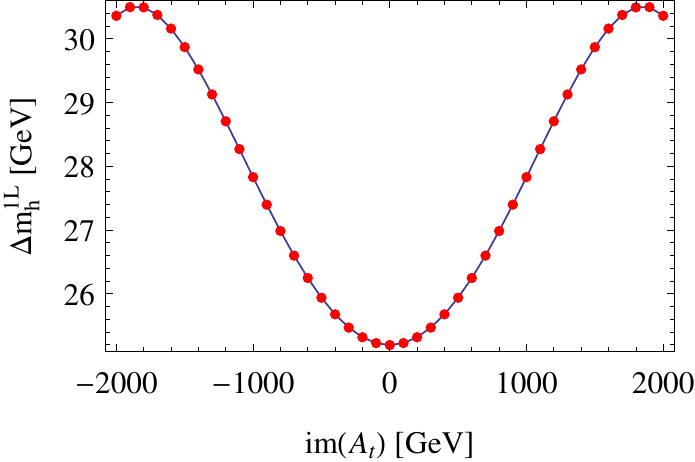}  \hfill
\includegraphics[width=0.32\linewidth]{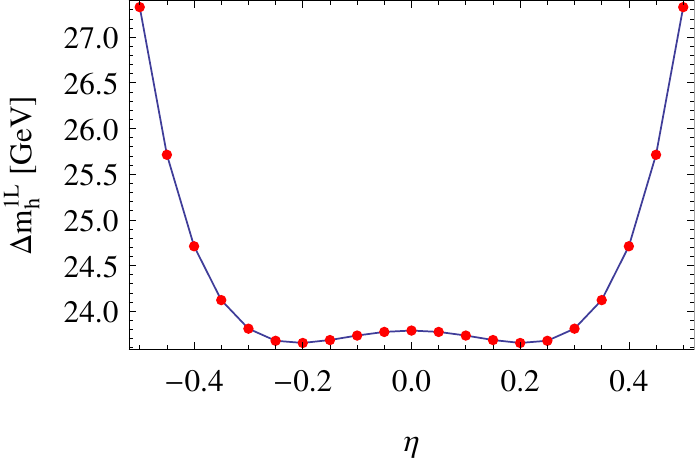}  \\
\includegraphics[width=0.32\linewidth]{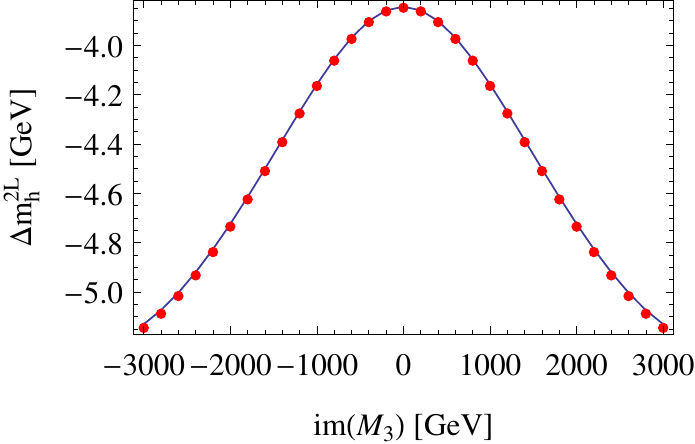}  \hfill 
\includegraphics[width=0.32\linewidth]{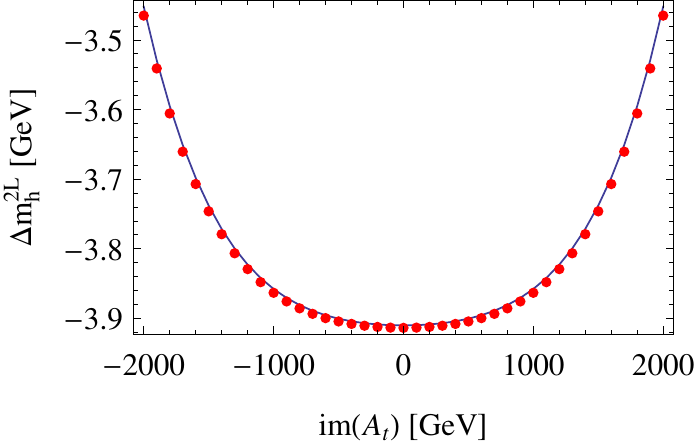} \hfill
\includegraphics[width=0.32\linewidth]{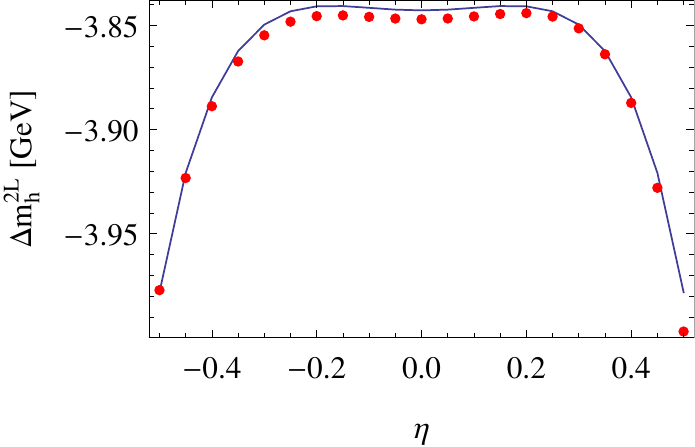}  
\caption{Effect of the one-loop corrections (left) and two-loop corrections (right) on the mass of the lightest scalars as function of 
im($M_3$) (first row),  im($A_{t}$) (second row) and $\phi_u$ (third row). The blue lines correspond to \SPheno, the red ones to \NC.  }
\label{fig:NMSSM3}
\end{figure}

The next check is to compare with existing results in literature. For the MSSM with CP violation the 
codes \FH \cite{Heinemeyer:1998yj} and \CPH \cite{Lee:2003nta} exist. However, both codes use another renormalisation scheme compared to \SPheno. Therefore,
already differences in the real case are present which are often larger than the expected effects from 
CP phases. Therefore, a quantitative comparison is not possible. The only other public code 
which supports CP violation is \NC for the (scale-invariant) NMSSM. 
\NC makes use of mixed $\overline{\mathrm{DR}}$--OS renormalization conditions for the computation of the Higgs
masses, but the OS effects in the (s)top sector can be turned off. This option together with some 
modifications described in the following allow for a very precise comparison. \\
The Higgs mass calculation at one-loop 
level is performed in \NC as in \SPheno including the full momentum dependence and all
possible contributions \cite{Ender:2011qh,Graf:2012hh}. At the
two-loop level only the ${\cal O}(\alpha_S \alpha_t)$ corrections are included
\cite{Muhlleitner:2014vsa}. The missing two-loop corrections will lead inevitable to a difference 
between \SPheno and \NC. Moreover, as has been discussed in detail in Ref.\cite{Staub:2015aea} the 
determination of the running \DRbar parameters entering the Higgs mass calculation also differs between 
both codes. Therefore, to have a meaningful comparison between codes in the case of CPV, we made the 
following modifications
\begin{itemize}
 \item {\tt \SARAH 4.8.3 and \SPheno 3.3.8}: 
 \begin{enumerate}
  \item All two-loop corrections but the ones ${\cal O}(\alpha_S \alpha_t)$ were turned off.
  \item The default input using SLHA-2 conventions \cite{Allanach:2008qq} were changed to SLHA-1 conventions \cite{Skands:2003cj} 
 for simpler comparison with \NC.
 \item The tadpole equations were modified to be solved for im$(A_\kappa)$ and im$(A_\lambda)$ instead of 
 im$(T_\kappa)$ and im$(T_\lambda)$ at tree-level: \NC solves the tree-level tadpole equations to calculate im$(A_\kappa)$ and im$(A_\lambda)$, but calculates the radiative shifts to 
im$(T_\kappa)$ and im$(T_\lambda)$, i.e. solves the loop-corrected tadpole equations with respect to other parameters than the tree-level
ones. The \SPheno code produced by \SARAH always solvs the tadpole equations at tree- and loop-level for the same parameters. This would have already given some difference  
at the one-loop level for specific complex phases, in particular for complex $\kappa$.
 \item The complex phases in the Yukawa couplings, which for instance appear via thresholds in the case 
 of complex $M_3$, were always put to zero becuase \NC supports only real 
 Yukawa couplings. 
 \end{enumerate}
 \item {\tt \NC 2.0}: 
 \begin{enumerate}
 \item  A flag to calculate only tree-level masses has been included. 
  \item The internal calculation of the running SM parameters has been overwritten. Instead, the 
 values are now read in from the input file.  This makes it possible to use exactly the same values 
 as \SPheno calculates.
 \item The finite shifts to $g_1$, $g_2$ and $v$ were put to zero to have a pure \DRbar renormalisation.
 \item We fixed a bug in the two-loop calculation which we found during our comparison.
\footnote{The expression for $\delta^{(2)} M^2_{H^+}$, which is used to express the shifts in the two-loop scalar mass 
matrix, contained a wrong prefactor for $\delta t_{\sigma_d}$.}
 \end{enumerate}
\end{itemize}
The conventions for the phases of the Higgs fields are 
\begin{align}
\label{eq:Hphases}
H_d \equiv \twov [\frac{1}{\sqrt{2}}(v_d +  \phi_d + i \sigma_d),H_d^-],\qquad H_u \equiv e^{i\eta} \twov[H_u^+,\frac{1}{\sqrt{2}}( v_u + \phi_u + i \sigma_u)],\qquad S \equiv e^{i\eta_S} \frac{1}{\sqrt{2}}( v_S + \phi_S + i \sigma_S).
\end{align}
where $\eta$ is used as input, and $\eta_S$ is calculated from the complex input of $\mu_{\rm eff}$ and $\lambda$ via
\begin{equation}
\eta_S = \text{arg}(\mu_{\rm eff}) - \text{arg}(\lambda)  \,.
\end{equation}
As default point we have chosen
\begin{eqnarray}
& \lambda=0.6 \,,\quad\kappa=-0.3\,,\quad\,A_\lambda=200~\GeV\,,\quad A_\kappa=1000~\GeV\,,\quad \tan\beta=3\,,\quad \mu_{\rm eff}=250~\GeV&\nonumber\\
\label{eq:DefaultNMSSM}
&M_1 = M_2 = 1000~\GeV \,,\quad M_3 = 1500~\GeV \,,\quad A_t = 1500~\GeV\,,\quad m_{\tilde t_L} = m_{\tilde t_R} = 1000~\GeV &
\end{eqnarray}
All other sfermion soft masses were put to $1.5$~TeV, and all other $A$-terms to zero.  \\
In Fig.~\ref{fig:NMSSMlam} we compare the radiative corrections to the three lightest scalars as function of im($\lambda$), while in Fig.~\ref{fig:NMSSM2}
the impact of the im($\mu_{\rm eff}$), and im($\kappa$) is shown. Finally, Fig.~\ref{fig:NMSSM3} depicts the dependence on im($M_3$), im($A_t$) and $\eta$. 
We find an overall very good agremment at the one- and two-loop level. The differences are at most $O(10~\text{MeV})$, which corresponds to  
the agreement obtained in Ref.~\cite{Staub:2015aea} for the real case, and does not increase in the presence of very large phases. Even if it is 
only possible to compare the corrections $O(\alpha_s \alpha_t)$ already the large majority of generic possible diagrams is covered. In particular 
all generic diagrams involving fermions are included, i.e. this confirms the correct treatment of the terms $V^{(2)}_{\ov{FF}S}$ as discussed
in sec.~\ref{SEC:METHOD}.

\section{MSSM}
\label{SEC:MSSM}
In \SARAH, we express the Higgs doublets in the MSSM in the usual form shown in eq.~(\ref{eq:Hphases}).
Since we are considering CP-violation, the $\mu$-term and holomorphic soft-breaking parameters are allowed to be complex, and in general all of the neutral scalars $\phi_{u,d}, \sigma_{u,d}$ mix, with one state yielding the Goldstone boson of the Z. We use the pure  \DRbar renormalisation scheme, and once we use the measured values of the lepton and quark masses, electroweak gauge couplings and weak mixing angle, Z boson mass, to determine the corresponding \DRbar quantities (Yukawa couplings, gauge couplings, electroweak expectation value), we still have a choice of parameters to be eliminated by the scalar tadpole equations. If we define $B_\mu \equiv e^{i\varphi_{B_\mu}} |B_\mu|$ these read
\begin{align}
0=& m_{H_d}^2 + \frac{1}{2} c_{2\beta}^2 M_Z^2 + |\mu|^2 - t_\beta |B_\mu|\cos (\eta + \varphi_{B_\mu}) + \frac{1}{v_d} \frac{\partial \Delta V}{\partial \phi_d} \\
0=& m_{H_u}^2 -\frac{1}{2} c_{2\beta}^2 M_Z^2 + |\mu|^2 - \frac{|B_\mu|}{t_\beta}\cos (\eta + \varphi_{B_\mu})  + \frac{1}{v_u} \frac{\partial \Delta V}{\partial \phi_u} \\
0=& \sin (\eta + \varphi_{B_\mu}) |B_\mu| + \frac{1}{v_u} \frac{\partial \Delta V}{\partial \sigma_d} =  \sin (\eta + \varphi_{B_\mu}) |B_\mu| + \frac{1}{v_d} \frac{\partial \Delta V}{\partial \sigma_u}.
\end{align}
The last two equations are not independent due to the gauge symmetries. Note also that $\eta $ and $\varphi_{B_\mu}$ are not independent: in the above they appear in the combination $\eta + \varphi_{B_\mu}$ so at tree level $\eta = - \varphi_{B_\mu}$. 

In the on shell scheme used by Refs.~\cite{Frank:2006yh,Hollik:2014wea,Hollik:2014bua}, the charged Higgs mass and $\mu$ are taken as input parameters; this is equivalent to specifying $|B_\mu|$ in the above equations and using the third tadpole equation to determine $\eta + \varphi_{B_\mu}$. This has the advantage of using a physical input, but the disadvantage of disguising potentially large tuning in the underlying values, particularly for small $\tan \beta$ where the loop corrections to the Higgs mass must be large, and typically lead to large corrections to the other Higgs masses too as we shall see in the following. In the \DRbar scheme $\eta$ (hence $\eta + \varphi_{B_\mu}$) is not a fundamental parameter, but rather something that should be derived. On the other hand, it appears in the Higgs couplings and would therefore be complicated to solve for, requiring a computationally-expensive iterative procedure and problems with non-zero goldstone boson masses (since we would be violating the condition that the shift is linear in a mass-squared parameter required in sec.~\ref{SEC:METHOD}. Instead, we fix $\eta$ and use the tadpole equations to determine $B_\mu$. 
Since the tadpole corrections are typically small compared to $B_\mu v_u$  this is a small adjustment and we can regard our choice of $\eta$ as being, to a good approximation, equivalent to minus the phase of $B_\mu$. For expediency due to the very strong constraints upon it we take $\eta=0$; we must then regard this as a tuning between $B_\mu$ and the other CP-violating phases in the theory for large values of $\varphi_{B_\mu}$.

There then remain two options: one conventional choice, as in the CP-conserving case, is to solve the tadpole equations for $|\mu|^2$ and $B_\mu$; this is appropriate when we have GUT-scale boundary conditions where we expect the other soft masses to unify (and we shall use this choice in subsection \ref{sec:MSSM_GUT}). For our study with low-energy boundary conditions in subsection \ref{sec:MSSM_Low} we shall choose to solve for $m_{H_d}^2, m_{H_u}^2$ and $\mathrm{Im}(B_\mu)$, taking $\mu$ and $\mathrm{Re}(B_\mu)$ as input parameters. This has the advantage that the tree-level heavy-Higgs masses are simply fixed by $\frac{|B_\mu|}{\sin \beta \cos\beta}$ so is closer to the on-shell interpretation; but as we shall see the loop corrections can be so large as to render a direct comparision of calculations in the two approaches impossible. In fact, this is further exacerbated since Refs.~\cite{Hollik:2014wea,Hollik:2014bua} use the \emph{charged} Higgs mass as the on-shell parameter, and we are only able (so far) to calculate the charged Higgs mass to one loop order compared to their two.

\subsection{One-loop masses and tadpoles}

The stop/top sector dominates the one-loop corrections as in the CP-conserving case. If we are only concerned with the lightest Higgs mass in the decoupling limit, then the leading corrections in the top Yukawa coupling $y_t$ in the effective potential approximation are
\begin{align}
(m_h^{\rm approx})^2 = M_Z^2 c_{2\beta}^2 + 
 \frac{3}{32\pi^2 v^2} \bigg[& s_{2\theta}^2 (m_{\tilde{t}_1}^2 - m_{\tilde{t}_2}^2)^2 + 8 m_t^2 \log  \frac{m_{\tilde{t}_1}^2m_{\tilde{t}_2}^2}{ m_t^4} \nn\\
& +  \frac{s_{2\theta}^2}{2} (m_{\tilde{t}_1}^2 - m_{\tilde{t}_2}^2) \bigg( 8 m_t^2 - s_{2\theta}^2 (m_{\tilde{t}_1}^2 + m_{\tilde{t}_2}^2) \bigg)\log  \frac{m_{\tilde{t}_1}^2}{m_{\tilde{t}_2}^2}\bigg]
\label{EQ:oneloopapprox}\end{align}
where $m_{\tilde{t}_{1,2}} $ are the masses of the stop eigenstates, $m_t$ is the top mass and the square of the sine of twice the mixing angle $\theta$ is defined as
\begin{align}
s_{2\theta}^2 \equiv& \frac{2 v^2 |e^{i\eta} T_{u}^{3,3} s_\beta - y_t \mu^* c_\beta|^2}{(m_{\tilde{t}_1}^2 - m_{\tilde{t}_2}^2)^2}.
\end{align}
Note that the stop mixing includes an additional phase (compared to the CP-even case) corresponding to the phase of $e^{i\eta} T_{u}^{3,3} s_\beta - y_t \mu^* c_\beta$, but we do not need that here.
The important observation here is that if we define $\mu \equiv e^{i \varphi_\mu }|\mu|, T_u^{3,3} \equiv e^{i\varphi_u}|T_u^{3,3}|$ then the phases only enter through the combination $\cos (\eta + \varphi_u + \varphi_\mu)$, and the result should be, to leading approximation, even in that combination of phases. Since our two-loop calculations are performed in the effective potential approach in the gaugeless limit, then this should also be true at two loops. 

The tadpole contribution from the stops is also important to our calculation, in particular the tadpole for the $\sigma_u,\sigma_d$ fields. We find that the one-loop stop contribution to these tadpoles neglecting the gauge couplings is
\begin{align}
\frac{\partial \Delta V}{\partial \sigma_u} \supset& - \frac{3v}{16\pi^2} \mathrm{Im}\bigg[T_u^{3,3} e^{i\eta} \big( e^{i\eta} T_u^{3,3} s_\beta - y_t \mu^* c_\beta \big)^* \bigg] \left(\frac{{\mathbf A}_0(m_{\tilde{t}_1}^2) - {\mathbf A}_0(m_{\tilde{t}_2}^2)}{ m_{\tilde{t}_1}^2 - m_{\tilde{t}_2}^2}\right) \nn\\
\frac{\partial \Delta V}{\partial \sigma_d} \supset& - \frac{3v}{16\pi^2} \mathrm{Im}\bigg[\mu^* y_t \big( e^{i\eta} T_u^{3,3} s_\beta - y_t \mu^* c_\beta \big)^* \bigg] \left(\frac{{\mathbf A}_0(m_{\tilde{t}_1}^2) - {\mathbf A}_0(m_{\tilde{t}_2}^2)}{ m_{\tilde{t}_1}^2 - m_{\tilde{t}_2}^2} \right)
\end{align} 
where we define ${\mathbf A}_0 (x) \equiv -x(\log x/Q^2 -1)$, $Q$ being the renormalisation scale. The function of masses on the right is a slowly-varying function with typical value of order unity, so with $\eta=0$ we have
\begin{align}
\mathrm{Im} (B_\mu) \sim &  \frac{3}{16\pi^2} | y_t \mu T_u^{3,3} | \sin \beta \sin (\varphi_u + \varphi_\mu) .
\end{align}  
So for  $\tan \beta = 5$ (for example) and $|T_u^{3,3}| = |\mu| = 2000\ \GeV, y_t \sim 0.9$ and maximal CP-violation in the combination of phases on the right hand side we have $\mathrm{Im} (B_\mu) \sim (270\ \GeV)^2 $. For a \emph{purely imaginary} $B_\mu$ this would correspond to tree-level charged/heavy Higgs mass of $600\ \GeV$; hence charged Higgs masses  below  $600\ \GeV$ -- or, more realistically, somewhat heavier --  invoke additional fine-tuning and are difficult to impose in the \DRbar scheme. In particular, this contributes to the fact that we cannot in any way reliably compare the results of our code to the benchmark scenarios of Refs.~\cite{Hollik:2014wea,Hollik:2014bua}, which involve lighter charged Higgs masses and larger trilinear couplings than we have quoted here.

\subsection{Alternative approach to the Higgs sector}

The MSSM is a special case as regards the two-loop mass computations in the gaugeless limit: there is no Goldstone Boson Catastrophe. To understand this, note that the tree-level Higgs potential consists only of the mass terms; the quartic couplings being given by the gauge couplings that are turned off. Hence the scalar masses are independent of the scalar expectation values in this limit. Now, the potential itself contains no divergences when taking the goldstone boson masses to zero -- the singluarities only appear in derivatives of the potential with respect to the goldstone boson masses -- and so the derivatives of the two-loop potential with respect to the scalar expectation values are finite. Hence we are free to consistently use the tree-level solution of the tadpole equations and mass matrices for the Higgs sector in the gaugeless limit, as was done for the calculations in the CP-conserving MSSM in Refs.~\cite{Degrassi:2001yf,Brignole:2001jy}. 

We can then write the neutral scalar mass matrix $\mathcal{M}^2_{h} $ and charged Higgs mass matrix $\mathcal{M}^2_{H^{\pm}}$ in the gaugeless limit in the basis $\phi_d, \phi_u, \sigma_d, \sigma_u$ as 
\begin{eqnarray}
& \mathcal{M}^2_{h} = \twoa[-\mathcal{M}_2^2 (-\tan \beta),0_{2\times2}][0_{2\times2},\mathcal{M}_2^2 (\tan \beta)] ,\quad\quad
\mathcal{M}^2_{H^{\pm}} = \mathcal{M}_2^2 (\tan \beta)  &
\end{eqnarray}
where
\begin{align}
\mathcal{M}_2^2 (x) \equiv& \twoa[|B_\mu| x, |B_\mu|][|B_\mu|,|B_\mu|/x].
\end{align}
Using these mass matrices in the diagrammatic two-loop routines thus neatly avoids tachyonic masses in the two-loop functions, although it does not significantly affect the result. 

\subsection{Two-loop Higgs mass with GUT boundary conditions}
\label{sec:MSSM_GUT}

If we take CMSSM boundary conditions, that is a unified A-term parameter $A_0$, scalar mass parameter $m_0$ and gaugino masses $M_{1/2}$, we should solve for $|\mu|^2$ and $B_\mu$ (both real and imaginary parts) at low energies (fixing the phase of $\mu$ as a choice).  Then due to the strong constraints from electric dipole moments the phases of $\eta, \mu, m_{12}$ are constrained to be very small (of order $10^{-3} \div 10^{-2}$) and thus not interesting parameters for the Higgs mass; we are only left with the phase of $A_0$. However, without performing a detailed scan to search for tuned corners of the parameter space, typically points that match LHC constraints on squarks and gluinos alongside reproducing the correct Higgs mass, will tend to have large values of $B_\mu$ and thus heavy additional Higgses, showing little CP-violating effects. 

\begin{figure}[hbt]
\centering
\begin{tabular}{cc}
\includegraphics[width=0.45\linewidth]{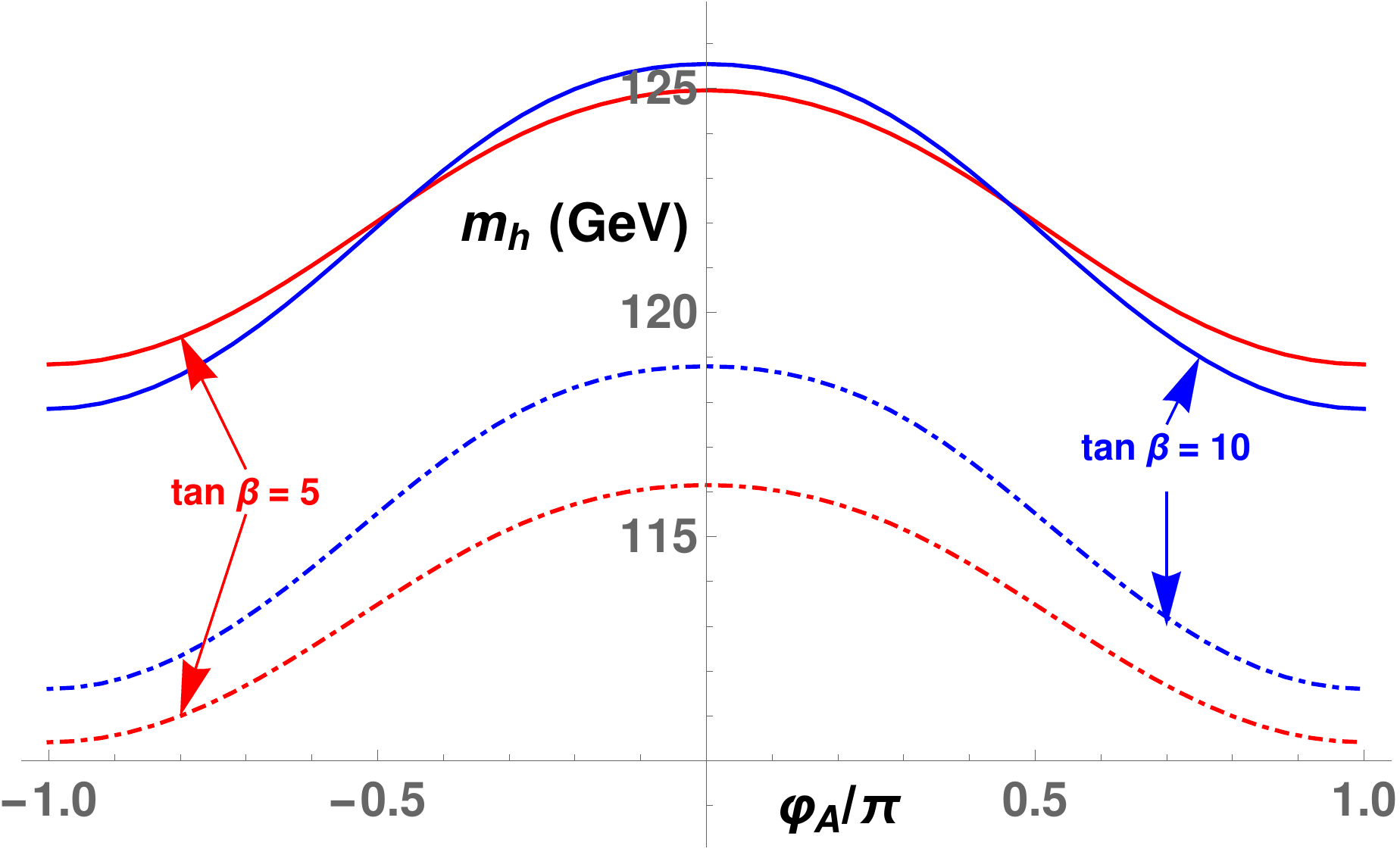} & \includegraphics[width=0.45\linewidth]{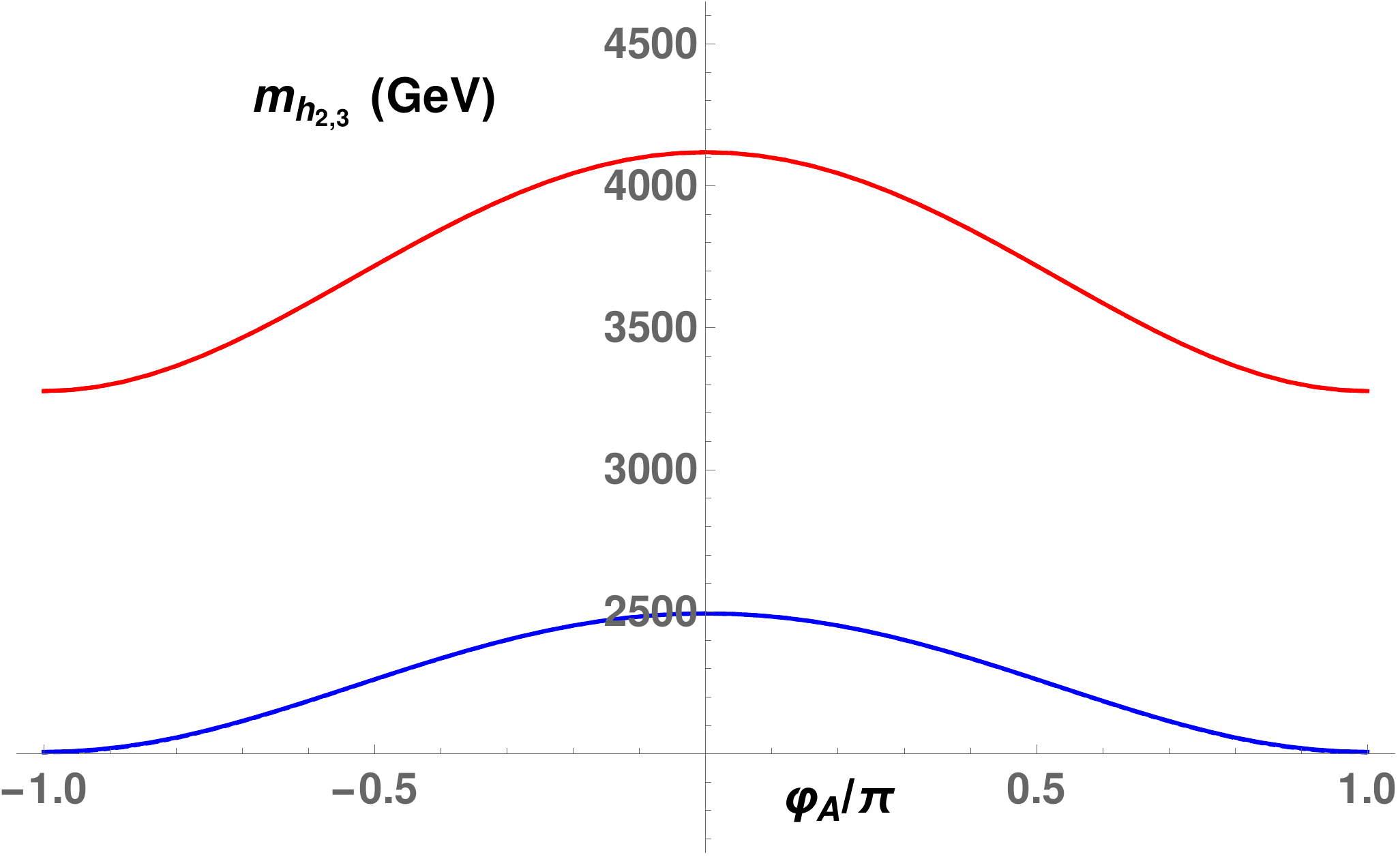} 
\end{tabular}
\caption{Plots of Higgs masses as the trilinear phase $\varphi_A$ is varied. On the left we show the lightest Higgs mass at one (dot-dashed) and two (solid) loops. On the right we show the masses of the heavier eigenstates which differ for each model by less than one GeV and so the distinction is not visible in the plot; the lower, blue, curve is for $\tan \beta =10$ while the upper, red, curve is for $\tan \beta = 5$.} 
\label{fig:MSSMGUT}
\end{figure}

We illustrate this in Fig.~\ref{fig:MSSMGUT}, where we define $A_0 \equiv -|A_0|e^{i\varphi_{A}}$ and vary the phase for two points:
\begin{center}
\begin{tabular}{||c|c|c|c||c|c|c||}  \hline 
 $\tan \beta$ & $|A_0|$\ (GeV) & $m_0$\ (GeV) & $M_{1/2}$\ (GeV) &  $m_{h_1}|_{\varphi_A =0}$\ (GeV) & $m_{h_{2}}|_{\varphi_A =0}$\ (GeV)& $m_{h_{3}}|_{\varphi_A =0}$\ (GeV) \\ \hline
 $5$ & $4500$ & $2000$ & $2000$ & $125.0$ & $4118$ & $4118$\\ \hline
 $10$ & $3300$ & $1500$ & $1000$  & $125.5$ & $2494$ & $2495$\\ \hline 
\end{tabular}
\end{center}
Here we have given the values that we compute at two loops for the three Higgs neutral Higgs scalars with $\varphi_A = 0$. In the figure we show the mass of the lightest Higgs at one and two loops as we vary $\varphi_A$; we also show the heavier Higgs masses whose differences are less than one GeV, either between one and two loops or between the second and third eigenstates. 

We selected small $\tan \beta$ values to maximise the visibility of the CP-violating effects, because in that way we can have a large difference between $\varphi_A =0$ and $\varphi_A = \pi$. This also requires large values of the trilinear couplings, and the contribution from stops to the Higgs mass must be large to obtain the correct value of the Higgs mass for at least some value of $\varphi_A$. However, this means that for typical points of the parameter space $B_\mu$ and $\mu$ will also be large, leading to heavy additional Higgses with small corrections to their masses at two loops. These states then have little impact on the light Higgs mass calculation and so the net effect is still little variation (of about $1$ GeV) in the two-loop contribution to the Higgs mass between $\varphi_A =0 $ and $\varphi_A = \pi$; almost all of the variation shown in the both plots of Fig.~\ref{fig:MSSMGUT} is due to the one-loop effects. 

\begin{figure}[hbt]
\centering
\includegraphics[width=0.6\linewidth]{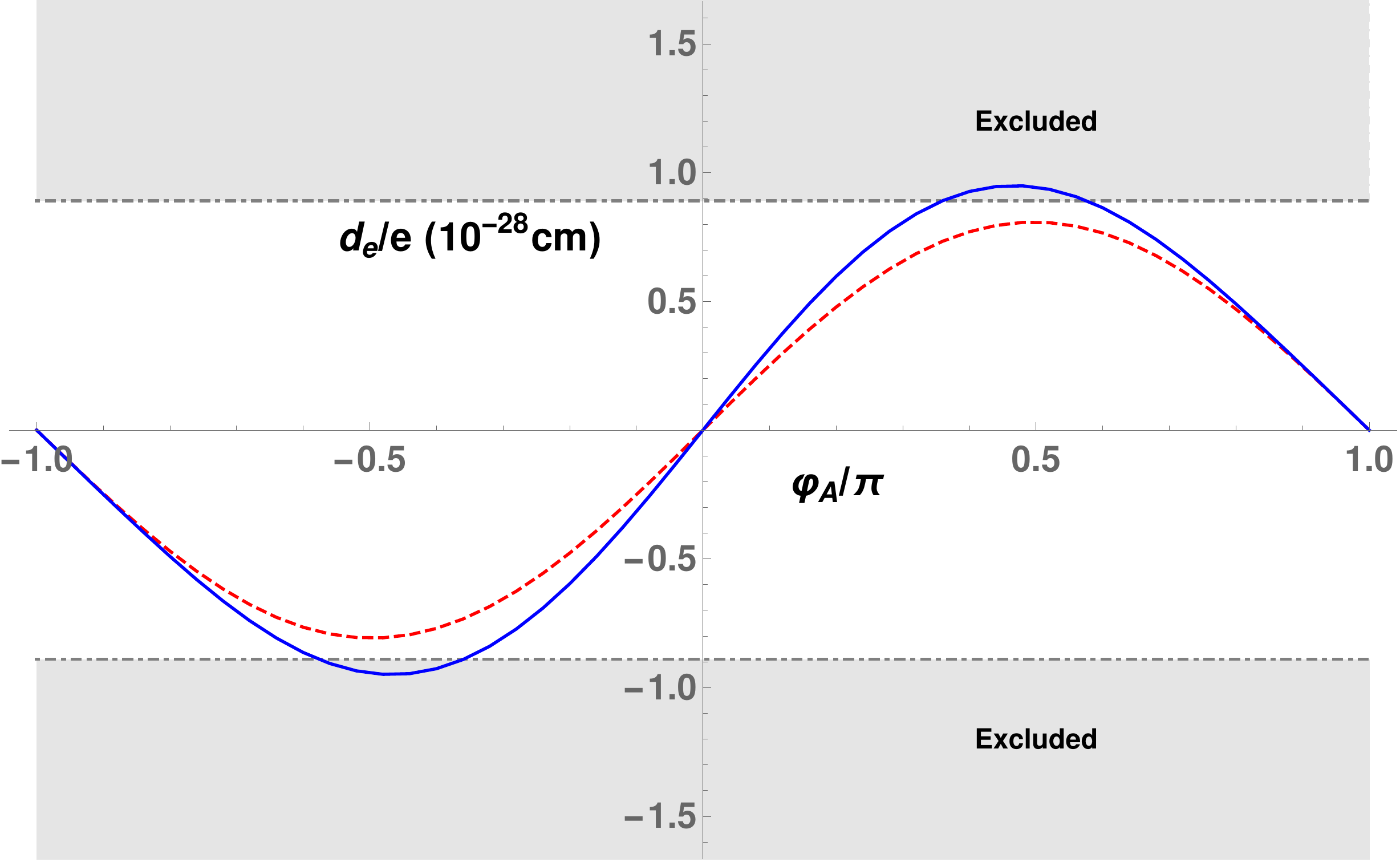} 
\caption{Electric dipole moment $d_e$ divided by electric charge $e$ as the trilinear phase $\varphi_A$ is varied, with exclusion bands shown, for values of $\tan \beta = 5$ (red) and $10$ (blue).} 
\label{fig:MSSMGUT_EDM}
\end{figure}

We finally note that the effect of large phases in the trilinears leads to gaugino phases through RGE running, and this in turn has a significant effect on the electron EDM $d_e$; we show the variation of this in Fig.~\ref{fig:MSSMGUT_EDM} and note that a small region near $\varphi_A = \pm \frac{\pi}{2}$ is already excluded for our $\tan \beta = 10$ scenario (recall that the contributions are proportional to $\tan \beta$ in e.g. eq.~(\ref{EQ:EDMchargino})). Thus we conclude that for CMSSM boundary conditions the CP-violating phases are well constrained and the specific two-loop corrections to the Higgs masses are less important. In the next subsection we shall consider more general low-energy boundary conditions. 

\subsection{Two-loop shifts with SUSY-scale boundary conditions}
\label{sec:MSSM_Low}

If we take our boundary conditions at low energies (i.e at the masses of the squarks and gauginos) then, without a particular bias for the conditions at high energies, we are free given the constraints to consider a phase of the trilinear terms and the gluino. Here we shall restrict our attention to the stop trilinear term $T_u^{3,3}$ and maximise the effect of CP-violation in the stop sector since this is typically the source of the largest contributions at two loops. Thus again we shall consider a small $\tan \beta$ scenario; we take as our parameter values
\begin{gather}
T_u^{3,3} = 0.9 A_0 e^{i\varphi_u},\quad T_d^{3,3} = 0.064 A_0 ,\quad T_e^{3,3} = 0.05 A_0 ,\nn\\
A_0 = 3800\ \GeV,\quad \mu=3800\ \GeV, \quad \tan \beta = 5,\quad \mathrm{Re}(B_\mu) = 10^6\ \GeV^2 \left(\frac{\tan \beta}{1 + \tan^2 \beta}\right)\nn\\
m_{\tilde{L},ii}^2=m_{\tilde{e},ii}^2= m_{\tilde{q},ii}^2 = m_{\tilde{u},ii}^2= m_{\tilde{d},ii}^2= (3\times 10^3\ \GeV)^2, \quad i=1,2 \nn\\
m_{\tilde{L},33}^2=m_{\tilde{e},33}^2= m_{\tilde{d},33}^2 = (10^3\ \GeV)^2,\quad m_{\tilde{q},33}^2 = (1.6\times 10^3\ \GeV)^2,\quad m_{\tilde{u},33}^2= (1.52\times 10^3\ \GeV)^2 \nn\\
M_1 = 200\ \GeV, \quad M_2=500\ \GeV,\quad M_3 = M_3^0 e^{i\varphi_{M_3}}.
\label{EQ:MSSM_LowPoint}\end{gather}
Here the prefactors of the trilinears are approximate values for the Yukawa couplings to allow simpler comparison between our points and those used in other works such as \cite{Hollik:2014wea,Hollik:2014bua}. The choice of the real part of $B_\mu$ (here we solve the tadpole equations for $m_{H_u}^2, m_{H_d}^2,\mathrm{Im}(B_\mu)$) is such that at tree level the heavy Higgs masses $m_{h_{2,3}}$ are at one \TeV.

\subsubsection{Variation of gluino phase}

\begin{figure}[!hbt]
\centering
\begin{tabular}{cc}
\includegraphics[width=0.45\linewidth]{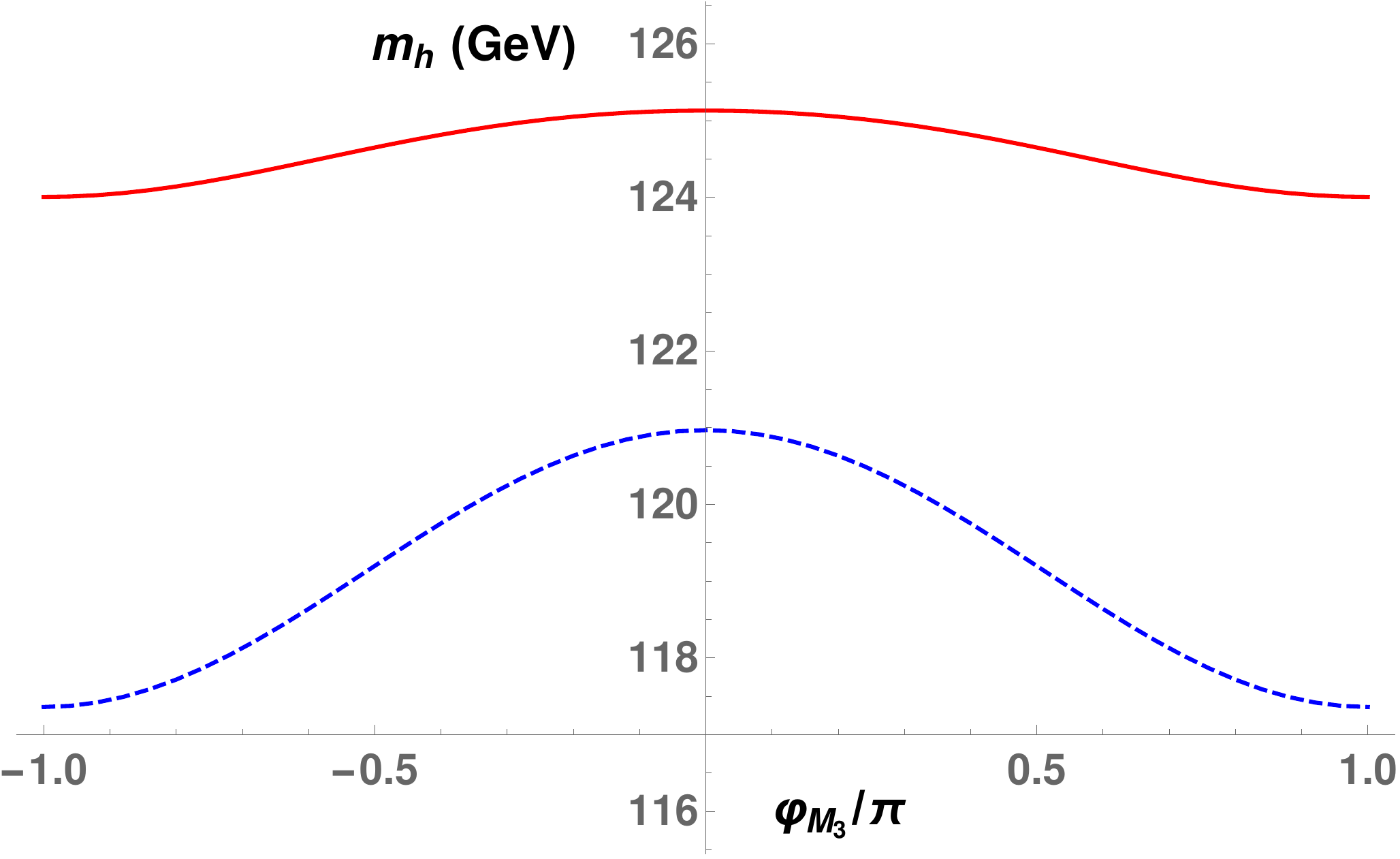} & \includegraphics[width=0.45\linewidth]{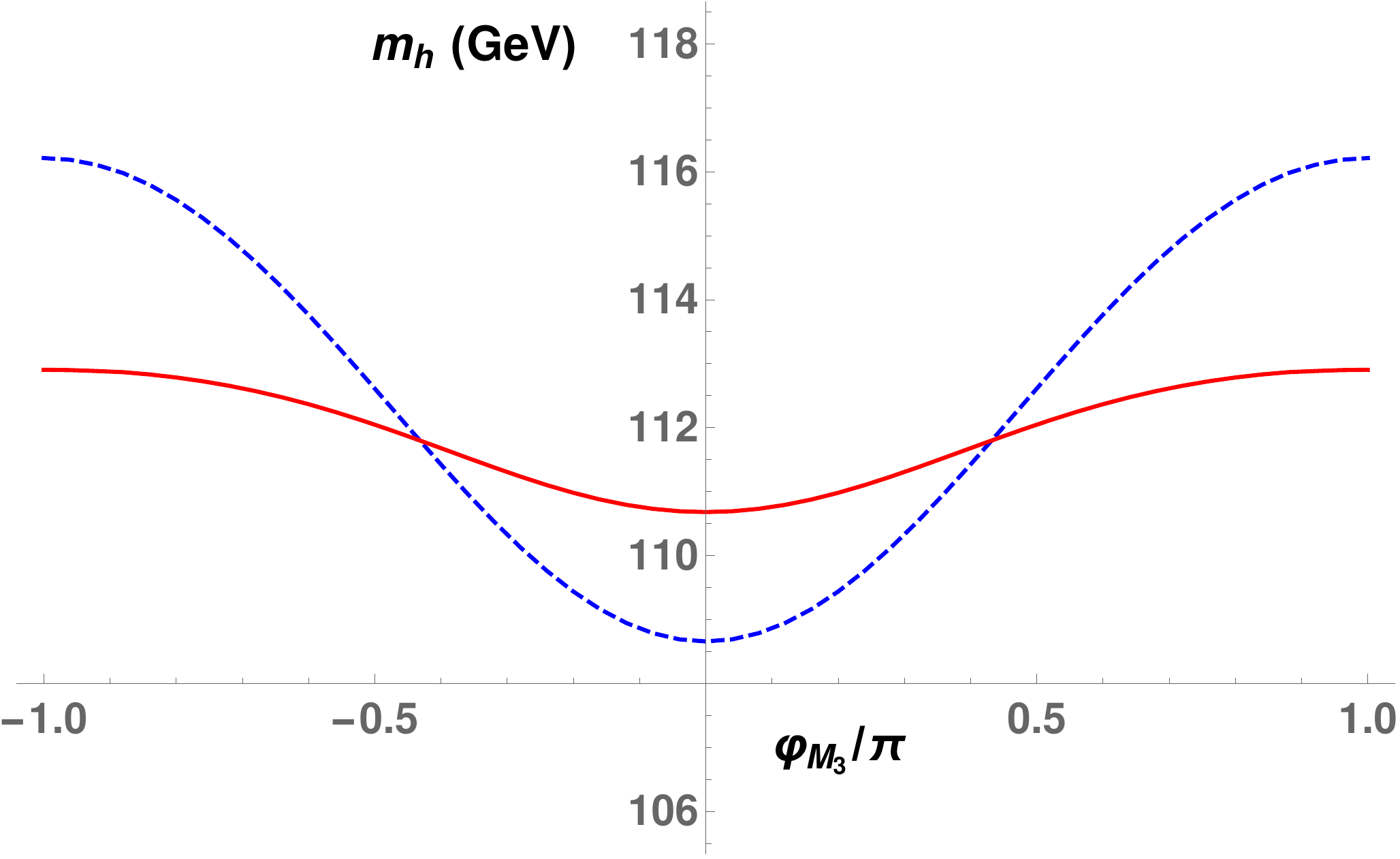} \\
\includegraphics[width=0.45\linewidth]{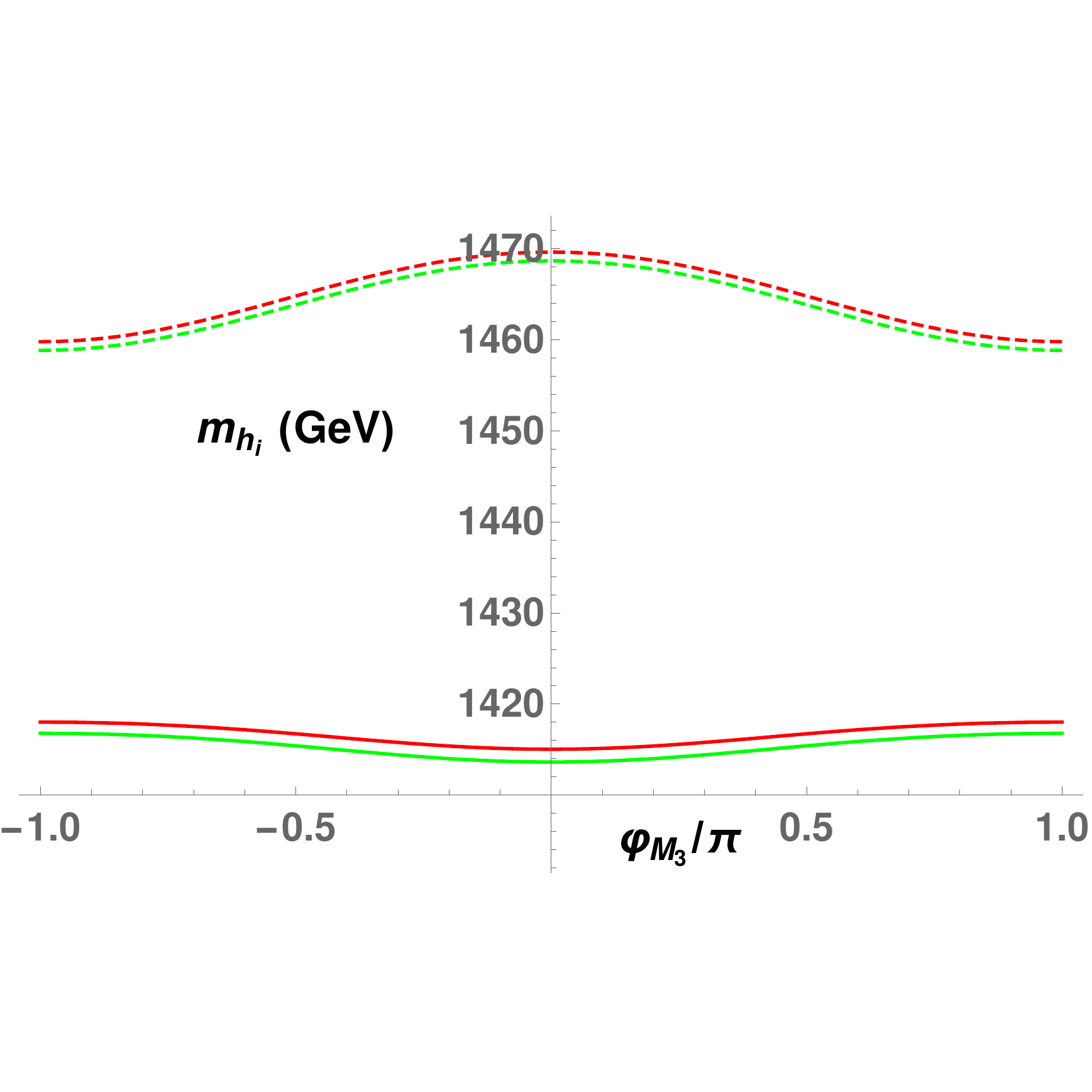} & \includegraphics[width=0.45\linewidth]{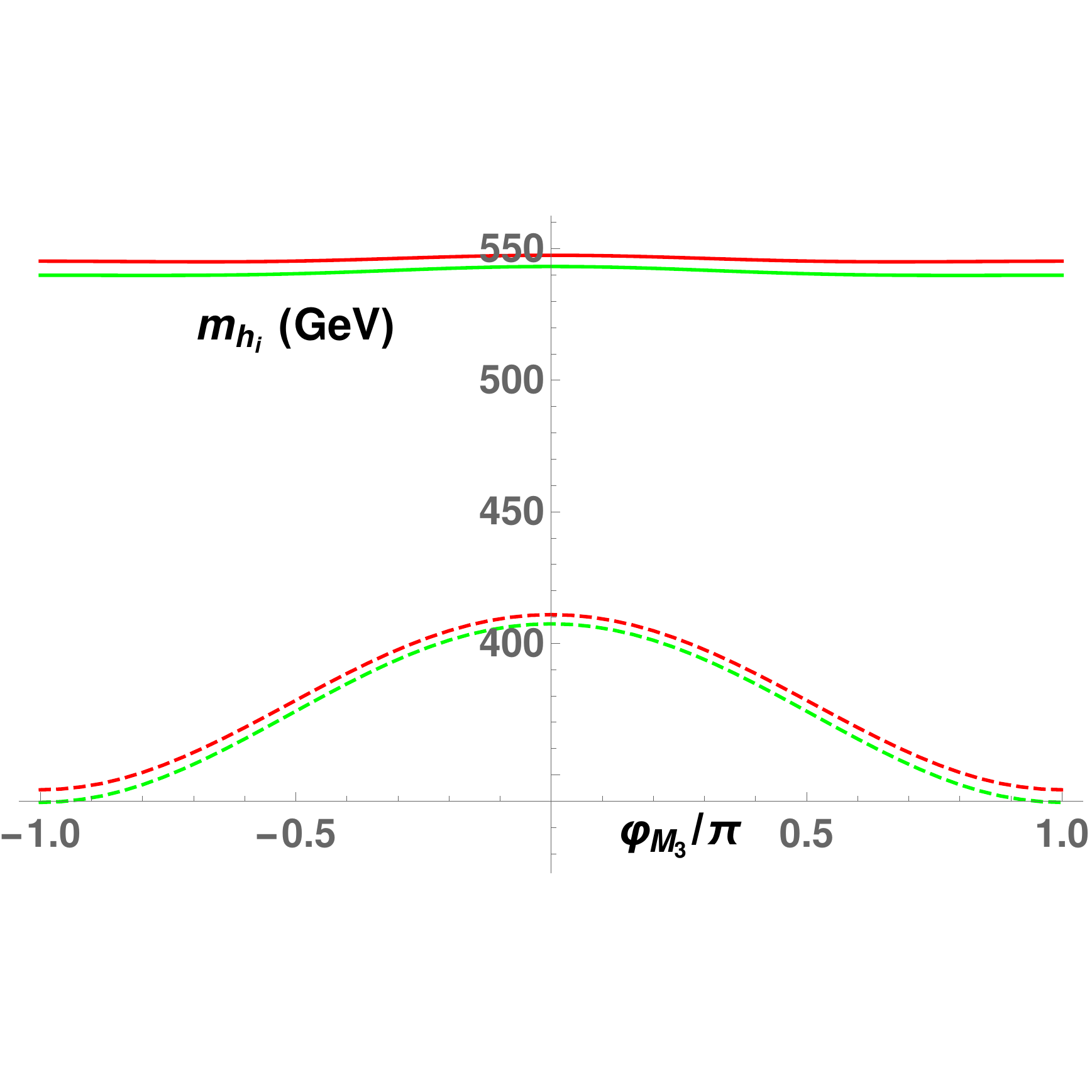} \\
\includegraphics[width=0.45\linewidth]{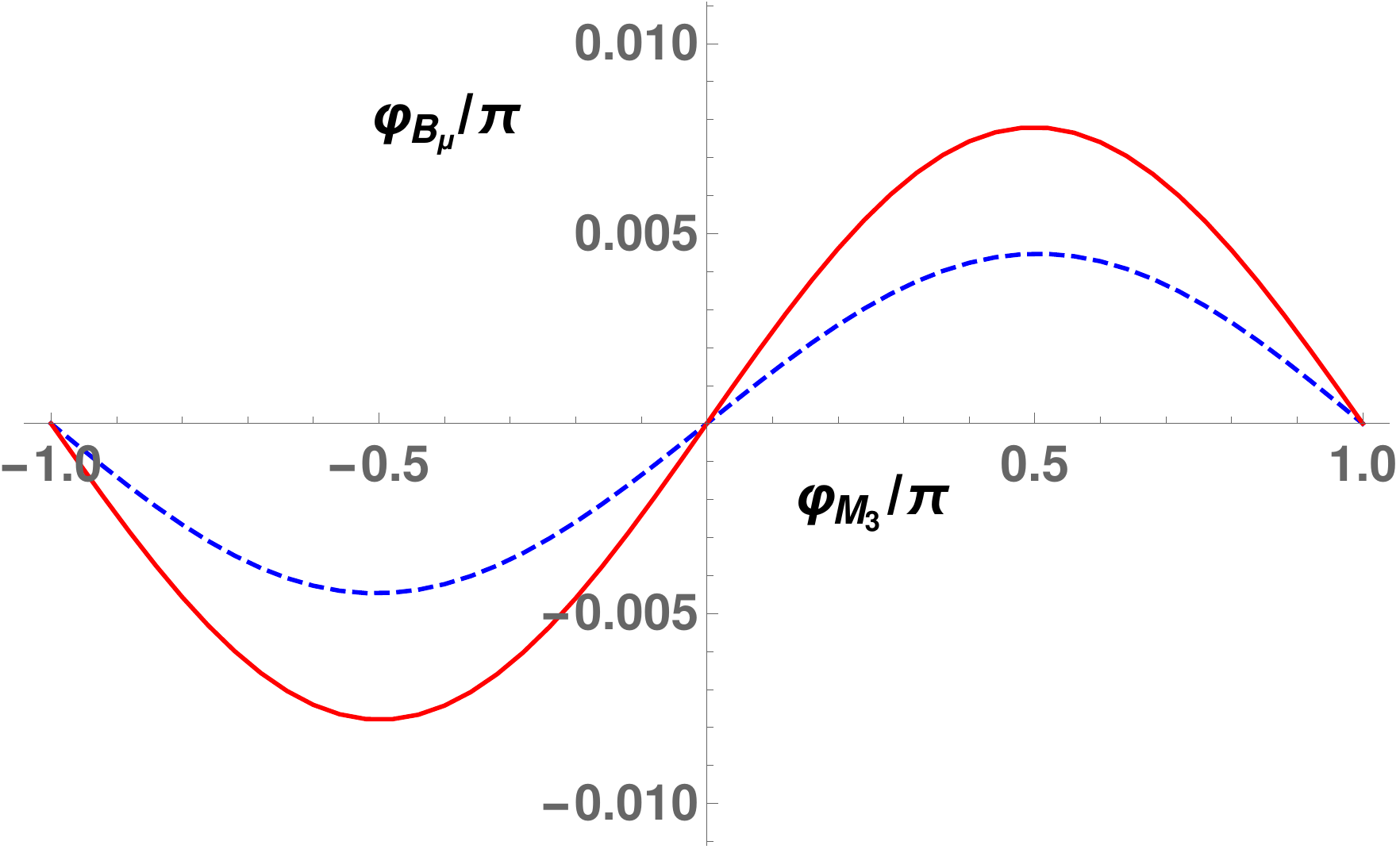} & \includegraphics[width=0.45\linewidth]{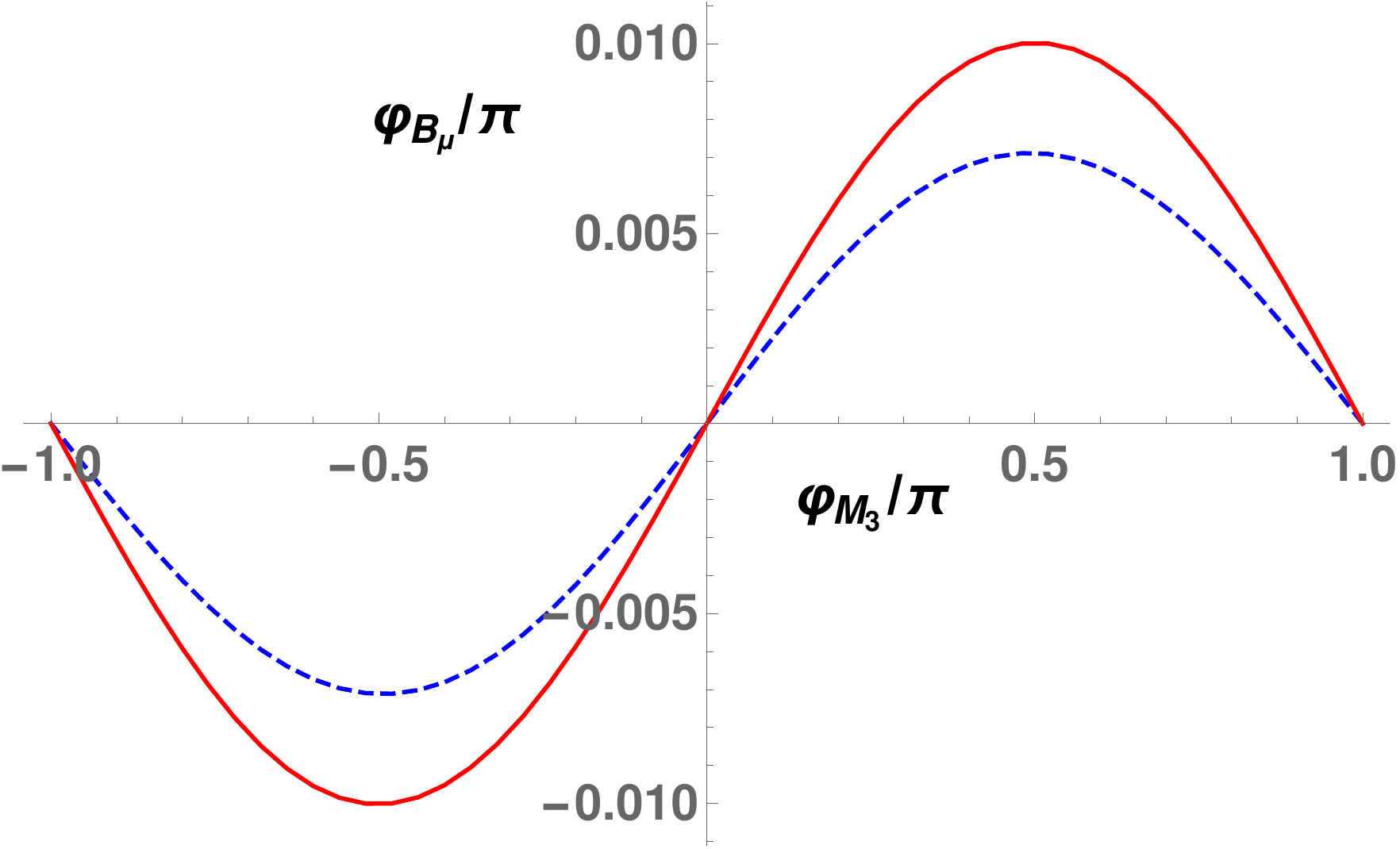}
\end{tabular}
\caption{Plots as the gluino phase $\varphi_{M_3}$ is varied for $M_3^0 = 2000$ \GeV; left plots have $\varphi_u = 0$, right plots have $\varphi_u = \pi$. The top plots show the lightest Higgs mass at one (blue, dashed) and two (red) loops; the middle plots show the next lightest (green) and heaviest (red) at one (dashed) and two (solid) loops. The bottom plots show the calculated variation of the phase $\varphi_{B_\mu}$ at one (blue, dashed) and two (red) loops.} 
\label{fig:MSSM_M3phase}
\end{figure}

We expect that the gaugino phase $\varphi_{M_3}$, only entering the Higgs mass calculation at two loops in the \DRbar scheme, should be an important parameter for our results. We show in Fig.~\ref{fig:MSSM_M3phase} the effect that it has on the three neutral Higgs masses and the parameter $\varphi_{B_\mu}$, for $\varphi_{u} = 0$ and $\varphi_u = \pi$. The first observation is that the difference between one and two loops is strongly dependent on $\varphi_{M_3}$; for $\varphi_u = 0$ this changes between $4$ and $7$ \GeV, and for $\varphi_u = \pi$ between $2$ and $-5 $ \GeV\ -- an overall shift of $7$ \GeV\ in the latter case! While this point has been chosen to show a large variation, it underlines the importance of the two-loop corrections.

Looking more closely, we find that the effect at two loops is partly to compensate for the variation at one loop, giving a more constant value for all Higgs fields. We also have the potentially counter-intuitive result that the $\varphi_u = \pi$ points have a \emph{smaller} Higgs mass than for $\varphi_u = 0$, when we might expect that when the $A$-terms are aligned with the $\mu$-terms we should have more mixing and thus larger masses. 

Both these observations have a simple explanation, that illustrates the need for the two loop routines. In the points that we have chosen with near maximal mixing, the soft terms are nearly degenerate, and so changes in the trilinear terms make only a small difference to the mixing angles. If we take the tree-level values of the stop masses and mixing (as we are required to do) and naively take $m_t = 173$ GeV, $v=246$ GeV we find for $\tan \beta = 5, M_3^0 = 2\TeV$ the following calculated values for  $m_h^{\rm approx} $ from eq.~(\ref{EQ:oneloopapprox}):
\begin{equation}
\begin{array}{||c|c|c|c|c|c||c||}\hline
\varphi_u & m_{\tilde{t}_1}\ (\GeV)& m_{\tilde{t}_2}\ (\GeV) & s_{2\theta}^2 & m_t\ (\GeV) & v\ (\GeV) & m_h^{\rm approx}\ (\GeV)\\\hline
0 & 1403 &1716 & 0.936 & 173 & 246 & 123 \\
\pi & 1320 & 1781 & 0.970 & 173 & 246 & 129 \\\hline
\end{array}
\end{equation} 
However, if we use the values actually calculated in \SPheno we find:
\begin{equation}
\begin{array}{||c|c|c|c|c|c|c||c||}\hline
\varphi_u & \varphi_{M_3} & m_{\tilde{t}_1}\ (\GeV)& m_{\tilde{t}_2}\ (\GeV) & s_{2\theta}^2 & m_t^{\DRbar}\ (\GeV) & v\ (\GeV) & m_h^{\rm approx}\ (\GeV) \\\hline
0 & 0 &1403 &1716 & 0.936 & 151 & 244.2 & 101 \\
\pi & 0& 1320 & 1781 & 0.970 & 145 & 243.6 & 88.3 \\
0 & \pi & 1403 &1716 & 0.936 & 147 & 244.6 & 95.6 \\
\pi & \pi & 1320 & 1781 & 0.970 & 152 & 243.0 & 100 \\\hline
\end{array}
\end{equation} 
Clearly the one-loop variation in the mass when we change $\varphi_{M_3}$ can only come from the change in the Yukawa coupling; the two loop shifts are then compensate for this (since the top-stop-gluino diagrams partly correspond to a self-energy correction to a top loop), which could presumably be more clearly seen in an on-shell scheme. What we also see is that when we vary $\varphi_u$ the shift in the top Yukawa has a much larger effect than the change in the mixing angle (since this can only be small when the mixing is already large). Therefore this observation is particular to the large mixing case; if the mixing were smaller, then potentially the variation of $\varphi_u$ could have the opposite effect and increase the Higgs mass as per our naive expectation.

The effect of $\varphi_u, \varphi_{M_3}$ on the heavy Higgses is very similar to the light Higgs: the loops compensate for the variation of the top Yukawa. But we note the enormous variation in their masses between $\varphi_u = 0, \pi$, and between one and two loops for $\varphi_{u} = \pi$; this underlines the tuning involved in the on-shell scheme when maintaining a constant heavy Higgs/charged Higgs mass. Finally, the variation of the phase $\varphi_{B_\mu}$ is significant enough that, if we were fixing the phase of $B_\mu$ and solving the tadpole equations for $\eta$, for most of the parameter space there would be large electric dipole moments; instead we find throughout that $|d_e/e| < 10^{-30}$.

\subsubsection{Variation of trilinear phase}

\begin{figure}[!hbt]
\centering
\begin{tabular}{cc}
\includegraphics[width=0.5\linewidth]{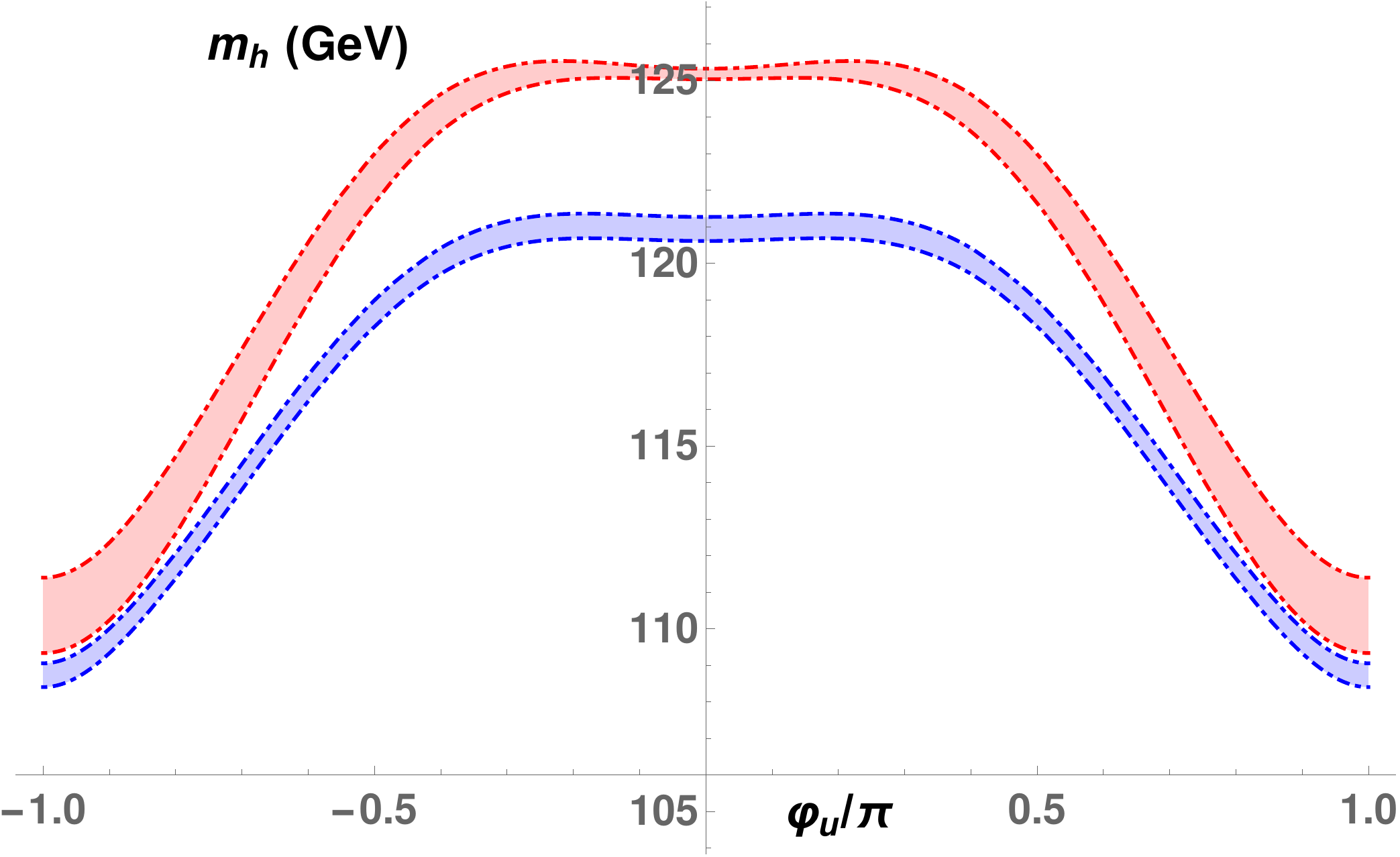} & \includegraphics[width=0.5\linewidth]{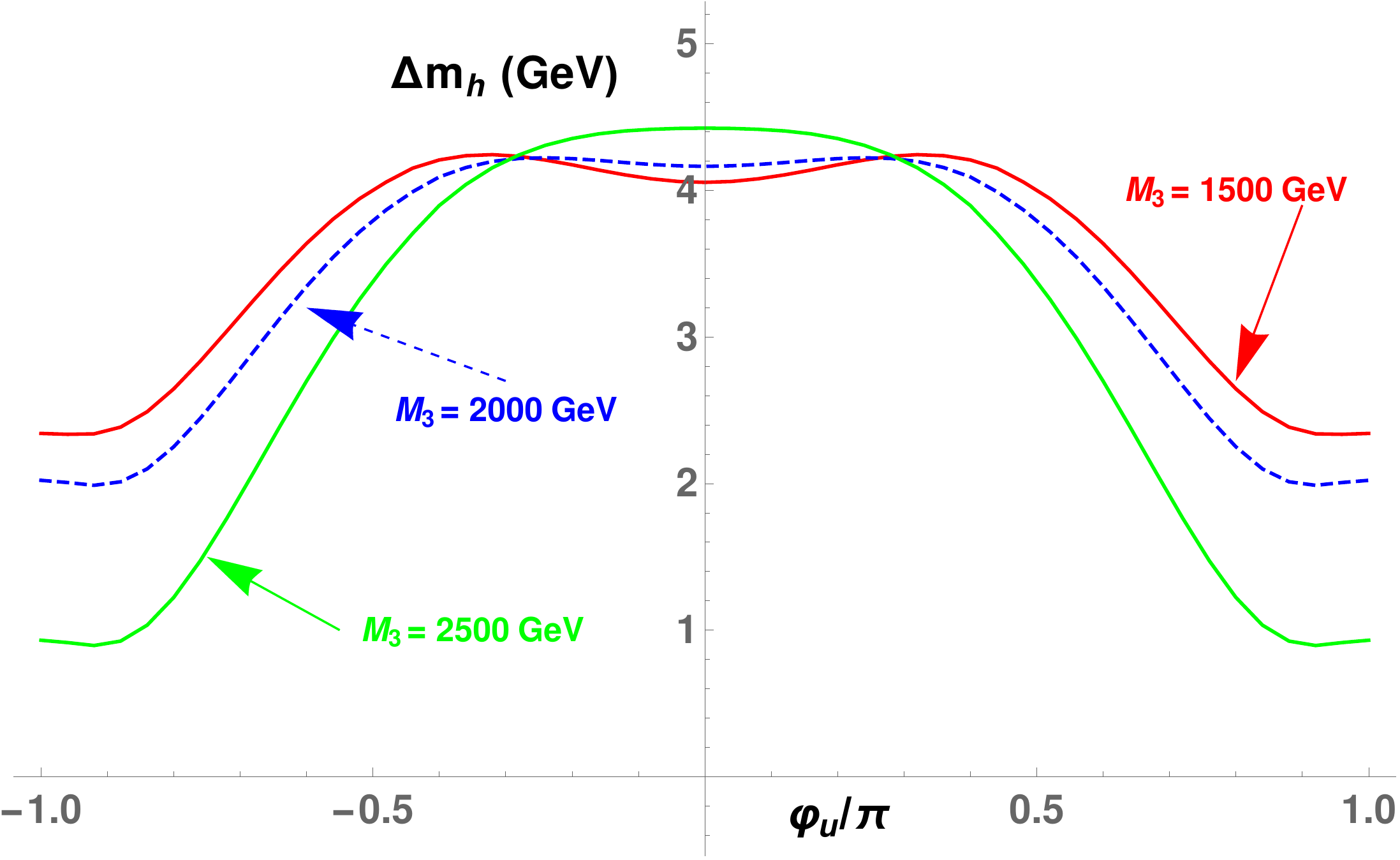} 
\end{tabular}
\caption{Left: plot of the lightest Higgs mass at one (blue) and two (red) loops as $\varphi_u$ is varied. The bands show the variation between $M_3 = 1500\ \GeV$ and $2500\ \GeV$. Right: difference between the one and two-loop lightest Higgs masses for three different values of $M_3$ marked in the plot, as $\varphi_u$ is varied.  } 
\label{fig:MSSM_atphase}
\end{figure}

\begin{figure}[!hbt]
\centering
\begin{tabular}{cc}
\includegraphics[width=0.5\linewidth]{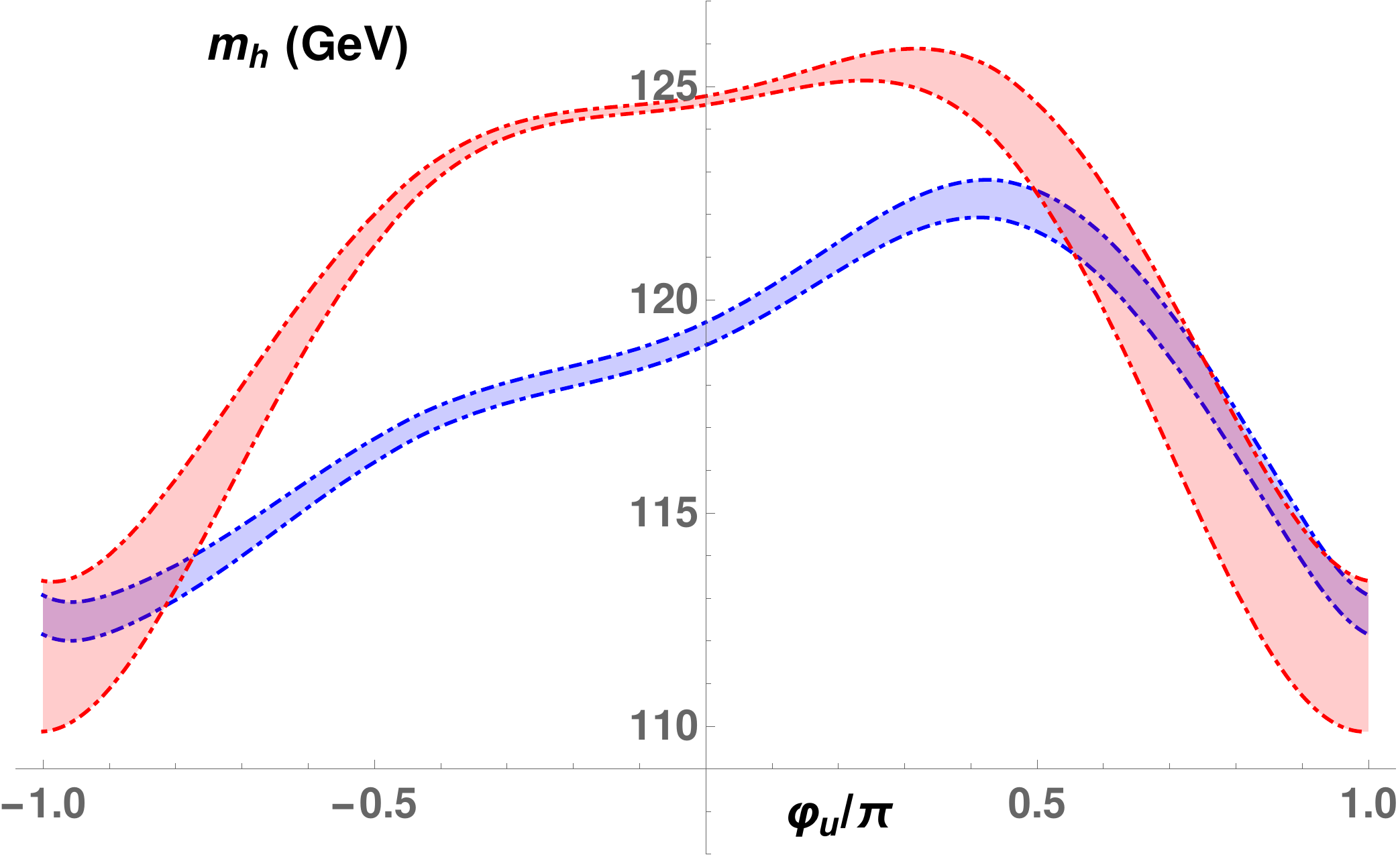} & \includegraphics[width=0.5\linewidth]{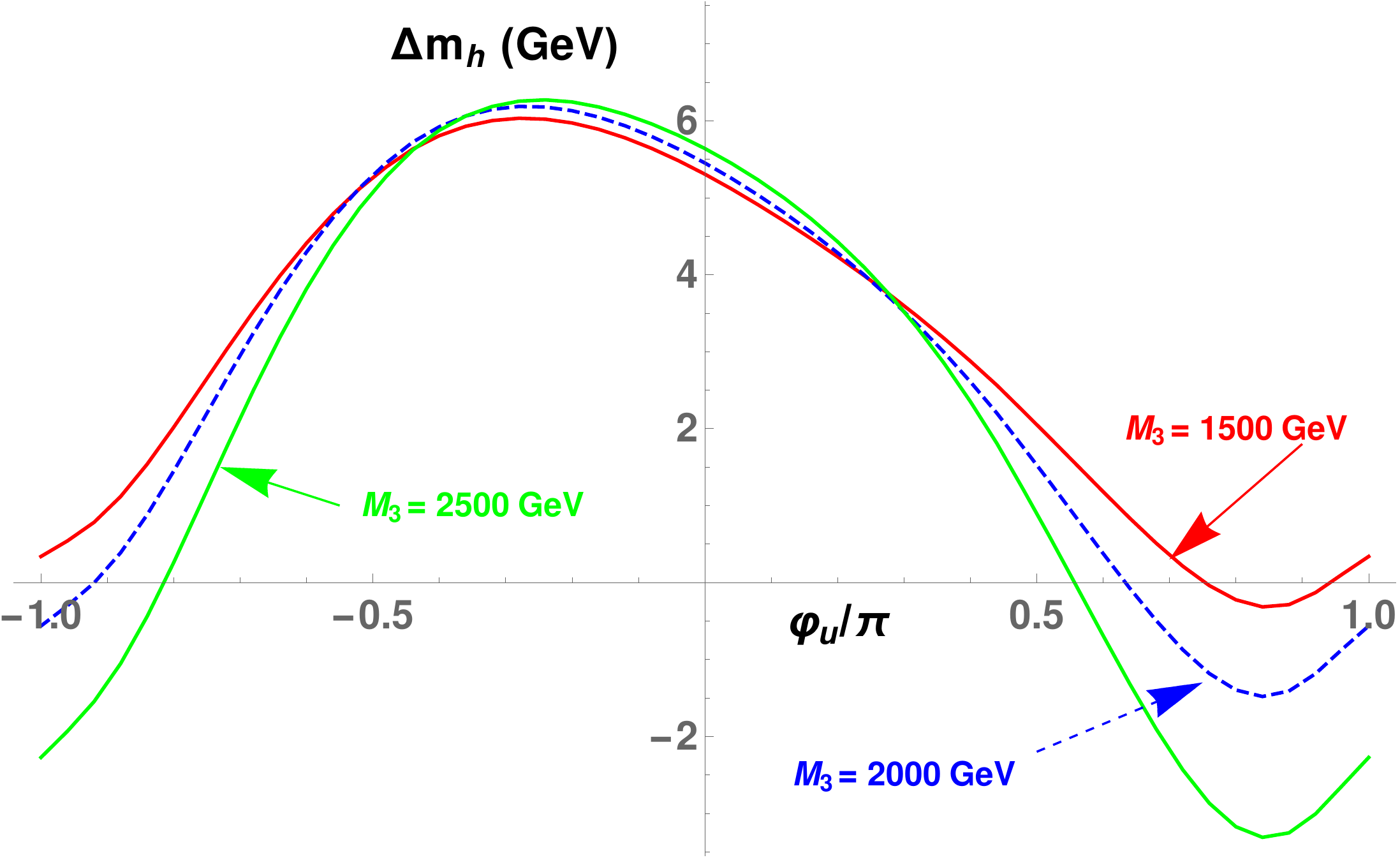} 
\end{tabular}
\caption{Same as figure \ref{fig:MSSM_atphase} but with $\varphi_{M_3} = \pi/2$.} 
\label{fig:MSSM_atphase_M3pio2}
\end{figure}

If we instead fix the gluino phase and vary $\varphi_u$, then we obtain Figs.~\ref{fig:MSSM_atphase} (for $\varphi_{M_3} = 0$) and \ref{fig:MSSM_atphase_M3pio2} (for $\varphi_{M_3} = \pi/2$). In those figures we show the variation of the light Higgs mass for three different values of $M_3^0$: the absolute value of the gluino mass clearly has a significant effect on the shift in the Higgs mass, of up to $2$ \GeV variation in the difference of one- and two-loop results by itself.

Overall we find that $\varphi_u$ has a markedly larger impact on the Higgs mass than $\varphi_{M_3}$; as we discussed in the previous subsection this is largely due to the shift in the top Yukawa coupling rather than the change in mixing of the stops. However, the effect of the two loop corrections as we vary $\varphi_u$ is clearly not to compensate for the shift in the top coupling: in contrast to the previous subsection we see large variations of $\Delta m_h$ (being the difference between two- and one-loop lightest Higgs masses) between $-2$ and $6$ \GeV\  in the maximally CP-violating case of $\varphi_{M_3} = \pi/2$; in that case we also see a significant asymmetry in the plot, which resembles in form figure 6 of Ref.~\cite{Hollik:2014bua} (indeed our parameter choices are deliberately similar), even though that calculation is in the on-shell scheme  and as we have already remarked cannot be quantitively compared. Even more than the previous subsection, this therefore underlines the importance of the two loop corrections to obtaining a realiable calculation of the Higgs mass. 
\section{Two-loop CPV effects in the NMSSM beyond $\mathcal{O}(\alpha_S\alpha_t)$}
\label{SEC:NMSSM}
\begin{figure}[hbt]
\centering
\includegraphics[width=0.5\linewidth]{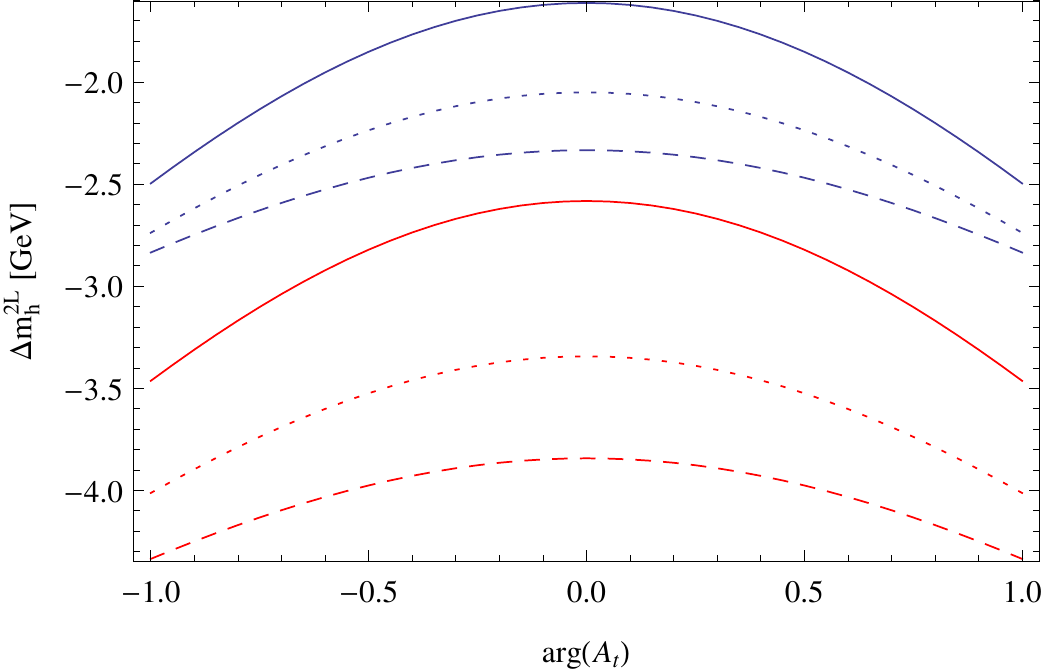}  
\caption{Size of the two-loop corrections as function of arg($A_t$) for different values of $|A_t|$: 1.5~TeV (dashed), 2.0~TeV (dotted), 2.5~TeV (full). 
The red lines include only corrections $\mathcal{O}(\alpha_S \alpha_t)$, while the blue lines include {\it all} other corrections in the gaugeless limit approximation.}
\label{fig:NMSSMdiffAtop}
\end{figure}
In sec.~\ref{SEC:VALIDATION}, we concentrated on the impact of complex parameters on the one-loop corrections as well as the two-loop corrections $\mathcal{O}(\alpha_S\alpha_t)$ in the complex NMSSM. However, with the combination \SARAH/\SPheno one can immediately go beyond this: all non-vanishing two-loop corrections in the gaugeless limit are included automatically. Therefore, we can check how well the entire impact of the complex phases is covered by  the $\mathcal{O}(\alpha_S\alpha_t)$ corrections. In the following, we use the same parameter values as in eq.~(\ref{eq:DefaultNMSSM}) if not stated otherwise. \\
We start with the phase of $A_t$ and show the change in the SM-like Higgs mass as function of arg($A_t$) in Fig.~\ref{fig:NMSSMdiffAtop} for three different values of $|A_t|$. As expected, the overall difference between the full two-loop corrections and the approximation using $\mathcal{O}(\alpha_S\alpha_t)$ grows with increasing $|A_t|$. However, for given $|A_t|$, the difference shows only a very mild depdence on the phase of $A_t$. Thus, at least for this parameter point the main sensitive of arg($A_t$) appears in the $(\alpha_S\alpha_t)$ corrections. 

\begin{figure}[tb]
\centering
\includegraphics[width=0.49\linewidth]{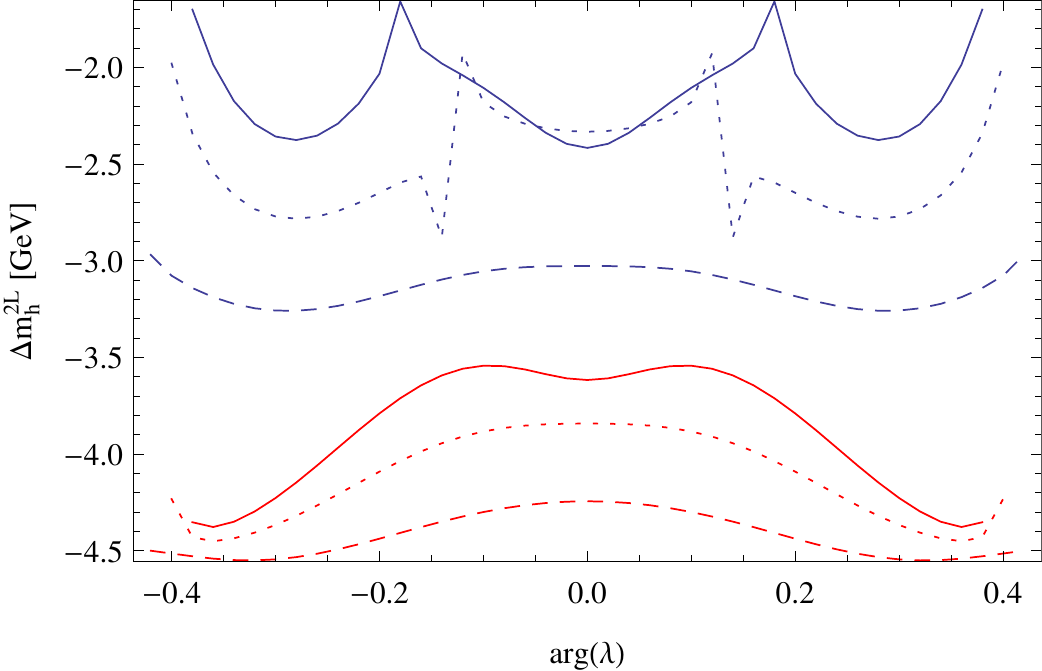} \hfill
\includegraphics[width=0.49\linewidth]{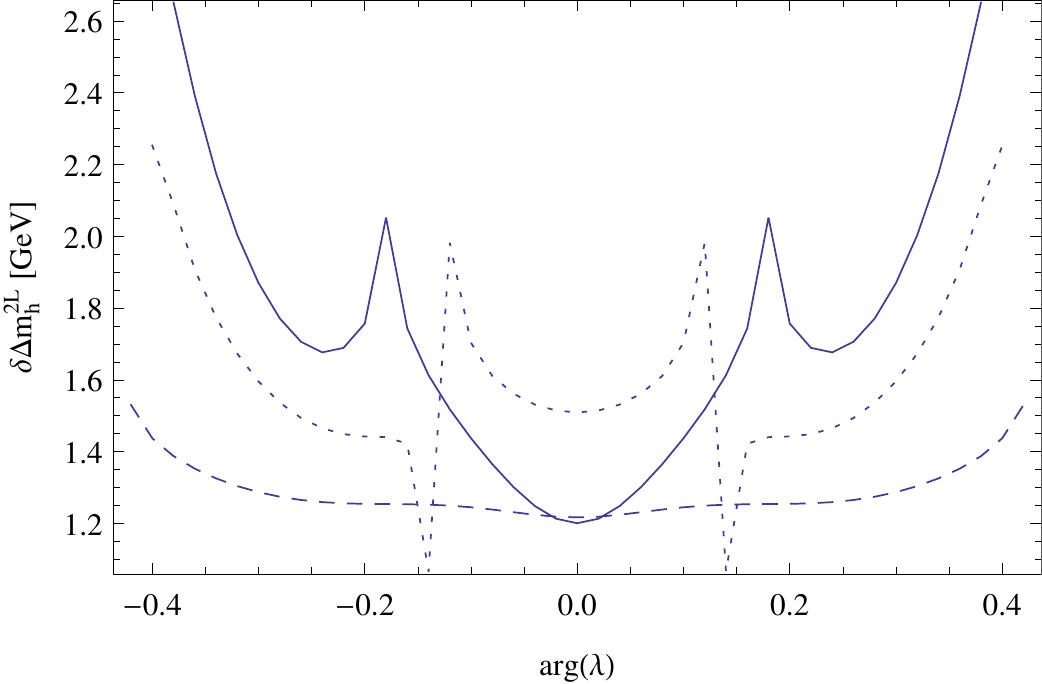} 
\caption{On the left: size of the two-loop corrections as function of arg($\lambda$) for different values of $|\lambda|$: 0.4 (dashed), 0.6 (dotted), 0.7 (full). 
The red lines include only corrections $\mathcal{O}(\alpha_S \alpha_t)$, while the blue lines include {\it all} other corrections in the gaugeless limit approximation. On the right: the differences between the red and blue lines. }
\label{fig:NMSSMdiffLambda}
\end{figure}

This is different for other phases like the one of $\lambda$: we show in Fig.~\ref{fig:NMSSMdiffLambda} the SM-like Higgs mass as function of arg($\lambda$) for three different values of $|\lambda|$. Here, we see not only a visible shift between the two different two-loop calculations, but also the dependence on the phase is very different. While the $\mathcal{O}(\alpha_S\alpha_t)$ corrections give the impression that the Higgs mass is reduced for a large phase of $\lambda$, the full calculation shows exactly the opposite. The Higgs mass actually increases for this point with increasing arg($\lambda$). As consequence, the Higgs mass is underestimated in the real case by the $\mathcal{O}(\alpha_S\alpha_t)$ corrections by about 1.3--1.6~GeV for all three vales of $|\lambda|$, while for arg($\lambda$) = $\pm$ 0.4 the discrepancy increases to 1.6--2.6~GeV. Thus, for singlet scenarios with large $\lambda$ and CP violation, the additional corrections now available with \SARAH/\SPheno can alter the SM-like Higgs mass by $O(\GeV)$. 

\begin{figure}[tb]
\centering
\includegraphics[width=0.49\linewidth]{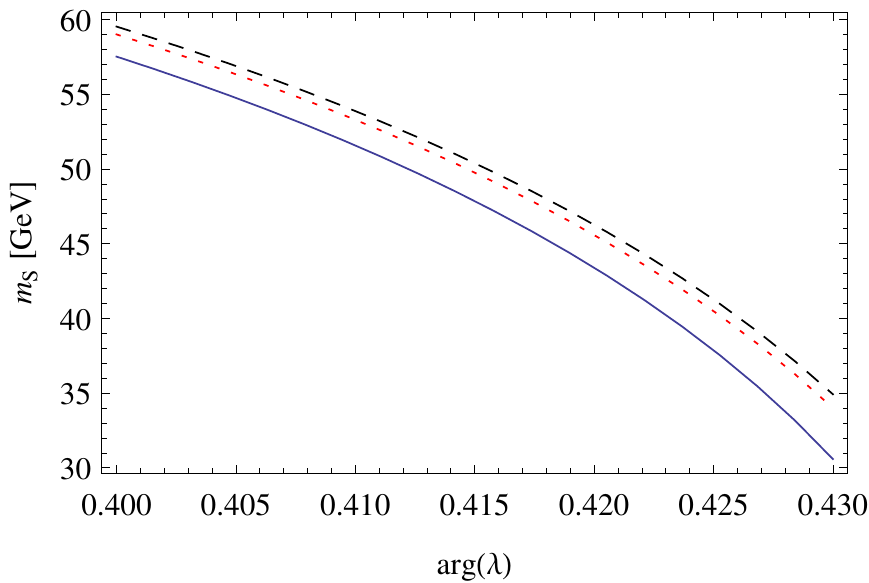} \hfill
\includegraphics[width=0.49\linewidth]{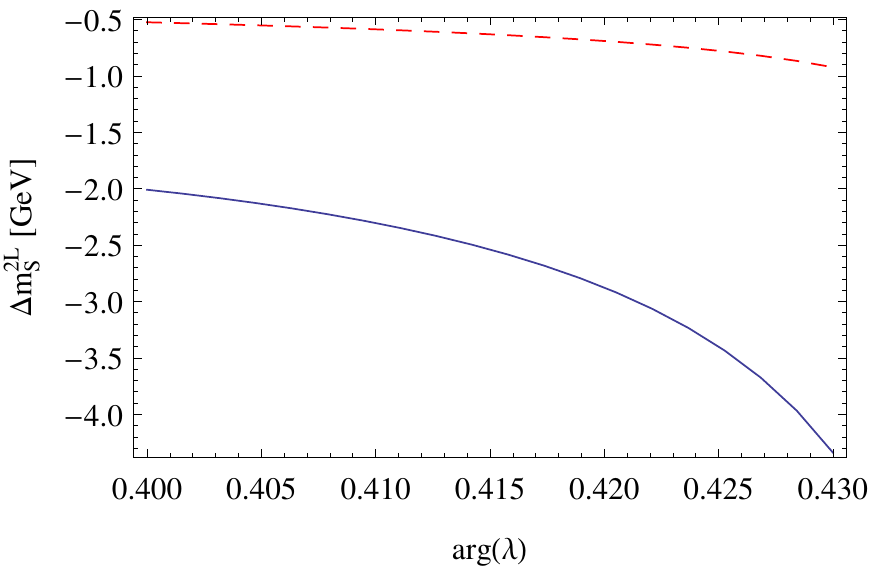} 
\caption{On the left: light singlet mass as function of arg($\lambda$) for the benchmark secenario TP3 of Ref.~\cite{Staub:2015aea}. The right side, show the size of the two-loop corrections $\mathcal{O}(\alpha_S\alpha_t)$ only (dashed, red line), and of the sum of all two-loop corrections (full, blue line). The black dashed line gives the mass at the one-loop level. }
\label{fig:NMSSMlightSingletLambda}
\end{figure}

\begin{figure}[tb]
\centering
\includegraphics[width=0.49\linewidth]{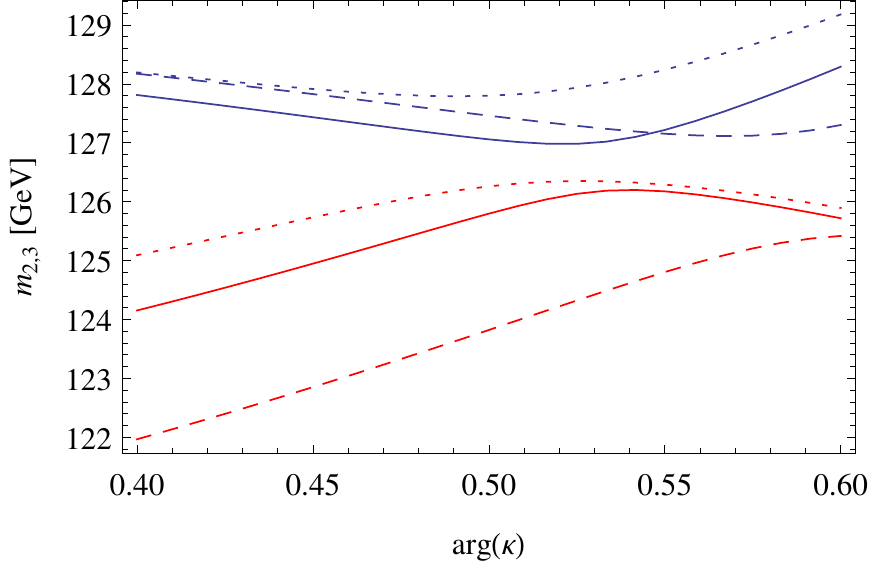} \hfill
\includegraphics[width=0.49\linewidth]{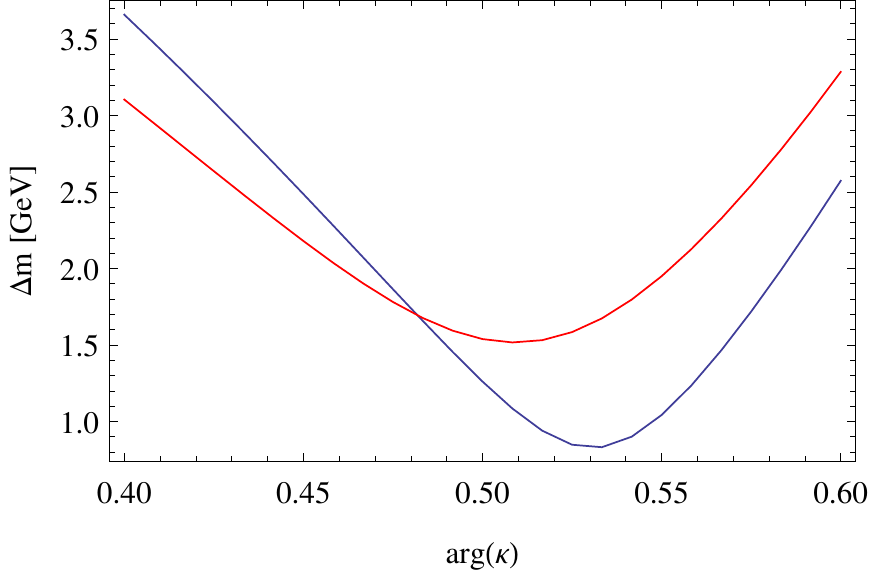}  \\
\includegraphics[width=0.49\linewidth]{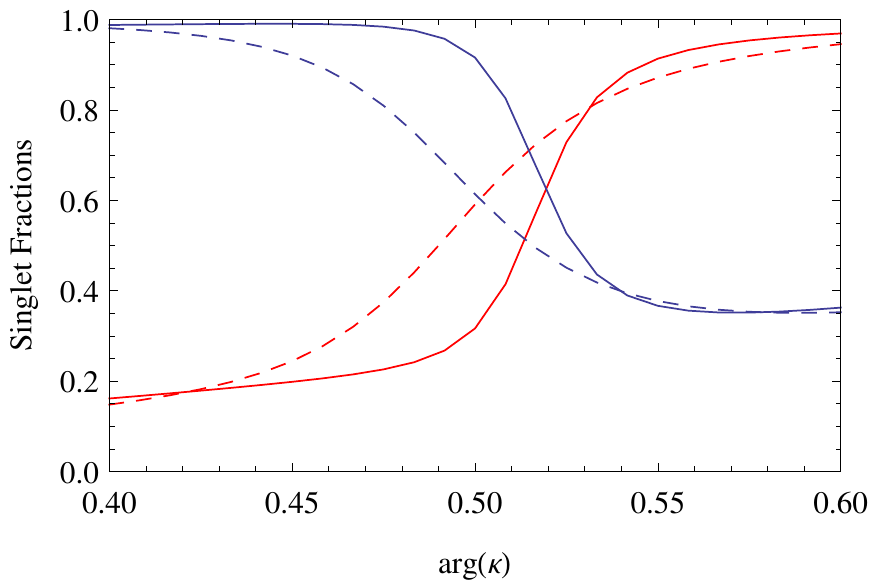} \hfill
\includegraphics[width=0.49\linewidth]{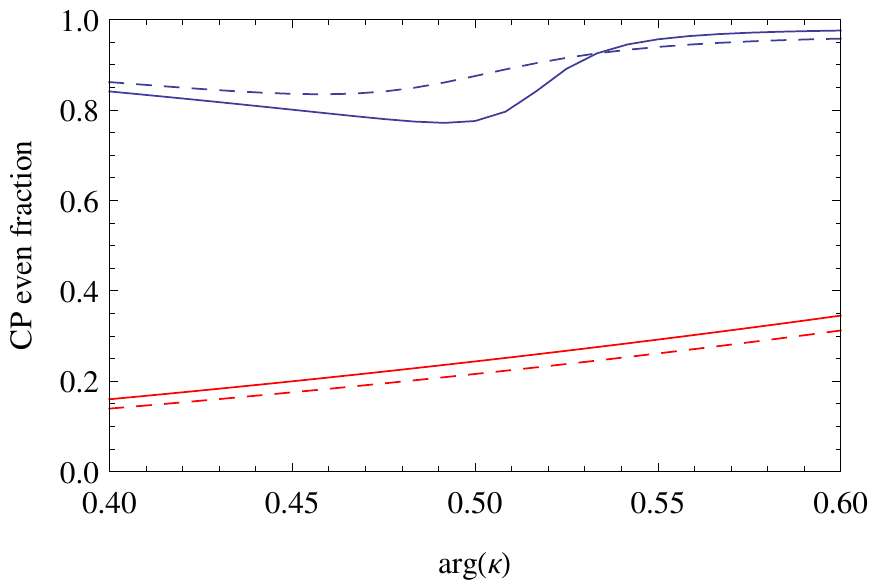}  
\caption{Impact of arg($\kappa$) for a scenario with two Higgs states close to 125~GeV as proposed in Ref.~\cite{Moretti:2015bua}. Top, left: The mass of the second and third physical scalar at one-loop (dashed), and including two-loop corrections $\mathcal{O}(\alpha_s \alpha_t)$ (dotted), as well as all two-loop corrections as calculated by \SARAH/\SPheno (full line). Top, right: difference between the scalars when using  $\mathcal{O}(\alpha_s \alpha_t)$ only (red), and the complete corrections (blue). Bottom, left: the singlet fraction of second (red) and third (blue) scalar using 2-loop corrections $\mathcal{O}(\alpha_s \alpha_t)$ only (dashed), and the complete 2-loop corrections in the gaugeless limit (full). Bottom, right: the CP-even fraction of the two states. The colour code is the same as on the left.   }
\label{fig:NMSSMkappa2H}
\end{figure}

Moreover, leaving the discussion of the mass of the SM-like Higgs for a moment, we find even bigger effects of arg($\lambda$) on the mass of a light singlet. This is depicted in Fig.~\ref{fig:NMSSMlightSingletLambda} where we have taken as basis the benchmark point TP3 of Ref.~\cite{Staub:2015aea}, but with a large phase arg($\lambda$)=0.40--0.43. In this range, the mass of the light singlet shows a large sensitivity to the phase of $\lambda$. We find here that the two-loop corrections  $\mathcal{O}(\alpha_S\alpha_t)$ shift the singlet mass between -0.5 and -1.0~GeV for this range, while the full two-loop corrections are about four times as large: for arg($\lambda$)=-0.4 they alter the singlet mass by -2~GeV, while for  arg($\lambda$)=-0.43 they even cause a shift of more than -4~GeV. 
We also briefly give an example for the effect of arg($\kappa$) by picking the very last point proposed in Tab. 3 of Ref.~\cite{Moretti:2015bua}. The particular feature of this point is that it has two scalars close to the desired mass of 125~GeV when choosing a phase of $\kappa$ of about 0.52. We show in Fig.~\ref{fig:NMSSMkappa2H} the sensitivity of the properties of these two scalars to changing this phase. First, we find that the actual phase at which the two states are closest in mass is only slightly different between the full two-loop calculation and the one including only corrections involving the strong interaction. However, the minimal difference in mass which we find for the full calculation is smaller than 1.0~GeV, while with the incomplete calculation it is not possible to come closer than 1.6~GeV when keeping all other parameters but the phase fixed. An even more important effect can be seen when considering the character of the two states: the singlet admixture of the doublet state is quite dependent on the used two-loop calculation. One finds for instance that the full calculation has a smaller mixing when moving away from the cross-over point than the $\mathcal{O}(\alpha_S\alpha_t)$ predicts. Also close to the cross-over point we find that the mixing between the CP-even and odd states is different between both calculations and the CP-odd component of the singlet would be underestimated by up to 10\% when not including all necessary two-loop corrections. 

\section{Conclusion}
\label{sec:conclusion}
We have presented the possibility of calculating the two-loop corrections to real scalar masses in SUSY models with CP violation using the public packages
\SARAH and \SPheno. After summarising the generic approach used in these calculations, we showed the self-consistency of all results and the perfect agreement 
with corrections implemented in \NC for the scale invariant NMSSM. We demonstrated for selected examples in the MSSM and NMSSM how interesting physical results can be obtained
easily with the available functionality, and that the variations of the corrections with the CP-violating phases can be large. We discussed the different options for the complex MSSM to fix CP phases by the tadpole equations, and the bias which enters the 
calculation by doing that. In the case of the MSSM, the only equivalent calculations have been performed in the on-shell scheme, rendering the results difficult to compare precisely; this underlines the utility of these results in \SARAH since the majority of spectrum generators deliberately use the \DRbar scheme for applications to studying GUT models, gauge-mediation, etc. On the other hand, we do find a pleasing qualitative agreement of our results.  

Afterwards, for the complex NMSSM we have briefly analysed the 
effect of CP phases in the two-loop corrections beyond $O(\alpha_s\alpha_t)$, which have not previously been available in any scheme. It was shown that the dependence on the phases of $M_3$ and $A_t$ is included to a large extent in the corrections involving the strong coupling. However, it turned out that for instance the effect of the phase of $\lambda$ in the full two-loop corrections deviates clearly from the impression one 
has when considering only the $\alpha_s\alpha_t$ corrections. 

Finally, we want to stress again that  it is now possible with the demonstrated approach to obtain the Higgs masses with very high accuracy not only for the CP-violating MSSM and NMSSM. A large variety of other SUSY models can now be easily studied in the presence  of significant CP phases without the problem of having a large theoretical uncertainty in the mass predictions. We hope that this gives a new impetus to interesting phenomenological studies of SUSY models with CP violation.

\section*{Acknowlegements}
We thank Dao Thi Nhung for clarifying discussions concerning \NC and Alexander Voigt for double checking one-loop results for the complex MSSM. MDG acknowledges support from Agence Nationale de Recherche grant
ANR-15-CE31-0002 ``HiggsAutomator,'' and would like to thank Pietro Slavich for interesting discussions.

\begin{appendix}
\end{appendix}

\bibliographystyle{ArXiv}
\bibliography{lit}

\end{document}